\documentclass[aps,twocolumn,superscriptaddress]{revtex4-1}

\usepackage{graphicx}
\usepackage{amsmath}
\usepackage{dutchcal}
\usepackage{bm}
\usepackage{color}

\usepackage[bookmarks=true,colorlinks=true,urlcolor=blue,linkcolor=blue,citecolor=blue,breaklinks]{hyperref}

\usepackage{amssymb}
\usepackage{booktabs}

\begin{document}

\sloppy

\newcommand{\blue}[1]{{\textcolor{blue} {#1}}}
\newcommand{\red}[1]{{\textcolor{red} {#1}}}
\newcommand{\green}[1]{{\textcolor{green} {#1}}}
%Title of paper
\title{Ultrafast light-driven optical rotation and hidden orders in bulk WSe$_2$}

\author{Emmanuele Cappelluti}
\email{emmanuele.cappelluti@ism.cnr.it}
\affiliation{Istituto di Struttura della Materia, CNR (CNR-ISM), 34149 Trieste, Italy}

\author{Habib Rostami}
\email{hr745@bath.ac.uk}
\affiliation{Department of Physics, University of Bath, Claverton Down, Bath BA2 7AY, United Kingdom}

\author{Federico Cilento}
\email{federico.cilento@elettra.eu}
\affiliation{Elettra-Sincrotrone Trieste S.C.p.A., 34149 Basovizza, Italy}
\maketitle

\textbf{
Single-layer semiconducting transition-metal dichalcogenides, lacking point inversion symmetry, provide an efficient platform for valleytronics, 
where the electronic, magnetic, valley and lattice degrees of freedom can be selectively manipulated by using polarized light.
This task is however thought to be limited in parent bulk compounds where the point inversion symmetry is restored.
Exploiting the underlying quantum physics in bulk materials is thus one of the biggest paradigmatic challenges.
Here we show that a sizable optical Kerr rotation can be efficiently generated in a wide energy range on ultrafast timescales
in bulk WSe$_2$, by means of circularly-polarized light.
We rationalize these findings as a result of the hidden spin/layer/valley quantum entanglement.
The spectral analysis reveals clear features at the three characteristic frequencies corresponding to the A-, B- and C-exciton edges.
The origin and the relative sign of all these features is shown to stem from the selective Pauli blocking of intralayer and interlayer optical transitions.
The long lifetime of the broadband Kerr response ($\tau \sim 500$ fs) provides a strong indication that coupled photo-induced electron and hole densities
survive in bulk compounds longer than previously expected.
The present report demonstrates that a hidden quantum entanglement is operative also in bulk centrosymmetric layered materials,
opening the way for an effective exploitation of bulk WSe$_2$ in optoelectronic applications. 
}

Semiconducting members of group-VI transition metal dichalcogenides (TMDs) stand out
as highly promising materials for the future of nanotechnology and quantum information \cite{liureview15,arora21,koperski17}.
These layered materials exhibit a wealth of unique physical properties when manipulated at the single-layer level.
Because of the bipartite honeycomb lattice structure and of the lack of inversion symmetry,
single-layer TMDs are characterized by
a new quantum number, the valley index, with a remarkable entanglement
with other inner degrees of freedom (charge, spin, orbital content, \ldots).
The valence bands at the two valleys $\pm$K are characterized by eigenstates
with opposite chiral structure, $|d_{x^2-y^2}\rangle\pm i |d_{xy}\rangle$, which results in a finite
Berry curvature with opposite sign in different valleys,
revealing a non-trivial topological character \cite{xiao12,liureview15}.
Furthermore, the chiral content of the valence bands leads
to different sensitivity of different valleys to the light polarization,
enforced in valley-dependent optical selection rules
\cite{xiao12,liureview15,arora21}.
All these features make TMDs an ideal platform for {\em valleytronics},
i.e.
the ability to selectively target a specific valley, and consequently a specific spin, using a tailored optical probe.
 \cite{soni22}.
This scenario has been widely explored in single-layer TMDs,
using a variety of experimental tools, as photoluminescence \cite{mak12nn,cao12,zeng12nn,kozawa14,lagarde14,lin21,lin22}, transport \cite{mak14science},
angle-resolved photoemission\cite{rostami19} and different optical and magneto-optical probes \cite{Wang_2013,Mai_2014,plechinger14,zhu14,dalconte15,yang15,yan17,plechinger17,mccormick18,kiemle20,arora21}.
A common feature of these techniques is the use,
as investigation tool,
the pump-driven dichroism  i.e. the difference of the probe signal between opposite circular polarization of the pump.
Bulk TMDs are generally considered less suitable for efficient manipulation of entangled quantum degrees of freedom.
Indeed, in bulk 2H-TMD compounds the unit cell contains two layers in an AB structure, which restores inversion symmetry. Consequently, under time-reversal symmetry, valley polarization is not permitted.

Despite this unpromising context, the potential to exploit similar tools in bulk materials for manipulating entangled degrees of freedom remains highly compelling, both from a theoretical perspective and for future market-oriented applications \cite{sun16}.
Recent research has shown that certain internal or {\em hidden pseudospin} degrees of freedom\cite{zhang14}
can indeed 
be harnessed in bulk materials  to restore functional properties
at the macroscopic level under equilibrium conditions \cite{gong13,zhang14}.
Such underlying hidden entanglement has been so far investigated
in TMDs mainly
by means of angle-resolved photoemission spectroscopy (ARPES),
with a specific focus on the
spin-polarization
of the photo-excited charges
in the conduction band,
with a spin-dependent
dichroism that disappears after very short timescales $t \sim 50-100$ fs \cite{riley14,bertoni16,beaulieu20,dong21,fanciulli23}.

Here we show that the optical response of centrosymmetric bulk WSe$_2$ can be efficiently manipulated in a wide energy range ($\hbar \omega \approx 1.4 -2.8$ eV), for a rather longer timescale of $\sim 0.5$ ps, by using a circularly polarized ultrashort pump pulse, with energy at the A-exciton frequency.
Such optical manipulation is revealed by a sizable transient optical Kerr rotation with clear features (and different signs) at the A- and B-exciton energies, as well as at the (high-lying) C-exciton.
So far, Kerr rotation 
was observed only in single-layer TMDs
with almost degenerate pump-probe setups,
both the pump and the probe energies
are set nearly
at the same A-exciton resonance.
We show that our results on bulk TMDs can be robustly rationalized as a consequence of the hidden entanglement between spin, valley, layer and orbital character, this latter ruled by the proper optical selection rules dictated by the Berry curvature. Using a microscopical quantum-field analysis
of the optical properties we correctly reproduce the experimental results and identify the origin and the relative signs of the three different features associated to the three different excitonic ranges.
The discovery of a relatively long lifetime for the pump-induced Kerr features suggests that 
in centrosymmetric bulk WSe$_2$
electron and hole excitations persist
in a bound state for significantly longer than previously indicated by ARPES data.
The present reports pave the way for an efficient optoelectronic manipulation of quantum degrees of freedom
in bulk TMDs, circumventing the limitations about growth and characterization of single-layer systems.

\section*{Ultrafast Kerr rotation revealed by transient dichroism}

\begin{figure*}[t]
\includegraphics [width=0.9\textwidth ]{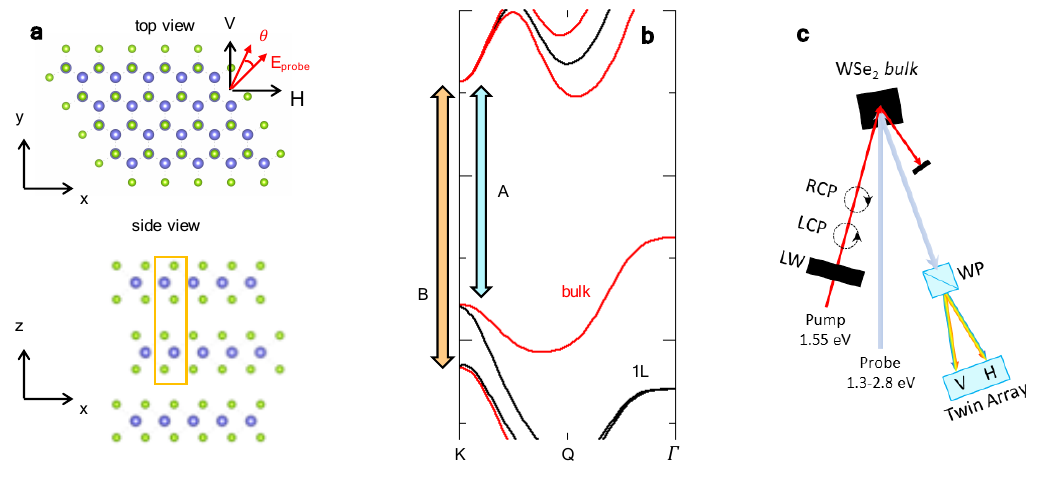}
\caption{
\textbf{Crystal structure, band structure and geometry of the time-resolved experiment.}
\textbf{a}, Crystal structure of the bulk WSe$_2$ crystal, from a top view (top panel) and a side view (bottom panel).
Tungsten (selenium) atoms are indicated in blue (green). The orange box denotes the bulk unit cell
including two WSe$_2$ layers.
Also shown here is the 45$^\circ$ polarization of the incoming probe photon,
whereas $\theta$ indicates the rotation angle of the polarization of the reflected probe beam.
\textbf{b}, Schematic band-structure of bulk TMDs (red lines), compared with the band structure of single-layer (black lines).
The vertical arrow represents the particle-hole optical transitions responsible for the A- and B-exciton resonances.
\textbf{c}, Sketch of the experimental configuration: the spectra of the horizontal (H) and vertical (V) components of a broadband supercontinuum beam ($1.3-2.8$ eV),
split and dispersed by a Wollaston polarizer (WP), are detected by a twin array detector. The polarization of the pump beam at
$\hbar\omega_{\rm pump}=1.55$ eV is switched from right-circular (RCP)
to left-circular (LCP) by means of a liquid crystal variable waveplate (LW). The probe impinges the sample at near-normal incidence, while the pump impinges
at an angle of $\approx$ 20$^{\circ}$. The bulk WSe$_2$ crystal is aligned as in panel (a), with its $x$-axis along the horizontal direction.
}
\label{fig:1}
\end{figure*}

The bulk semiconducting TMDs with chemical formula $MX_2$ ($M={\rm Mo}, {\rm W}$; 
$X={\rm S}, {\rm Se}$) share a 2H lattice structure where the unit cell contains two stacked layers $MX_2$ with a relative rotation of 180 degrees  (Fig.~\ref{fig:1}a).
It is known that the interlayer coupling in bulk TMDs changes many of these compounds from direct- to indirect-bandgap semiconductors,
with an indirect gap between the $\Gamma$ and Q points of the Brillouin zone (Fig.~\ref{fig:1}b) \cite{mak10,splendiani10}.
Despite of this, however, the dominant optical features of bulk systems stem from particle-hole excitations close to the 
$\pm$K points, just as for single-layer compounds, and are characterized
by two exciton features, termed A and B,
which are related, respectively, to the particle-hole excitations
between the spin-split upper and lower valence bands and the nearly-degenerate conduction band  \cite{mak10}.
Typical values of A and B exciton resonances in bulk WSe$_2$ at room (low) temperature
are $E_{\rm A} \sim 1.6 (1.7)$ eV, $E_{\rm B} \sim 2.2 (2.1)$ eV, respectively,
slightly larger than in single-layer compounds \cite{li14,arora15,morozov15}.
Like other TMDs, a further high-energy broad shoulder, usually denoted as ``C-exciton'', is reported in WSe$_2$,
which is often associated in literature with band-nesting conditions occurring along the $\Gamma$-K path
of the Brillouin zone.

In our setup\cite{perlangeli20},
in order to reveal in an efficient way the possible optical Kerr rotation,
we perform an experiment (see sketch in Fig.~\ref{fig:1}c) where a polarization-resolved broadband probe, covering the photon energy range $1.3-2.8$ eV, is used to detect
the pump-induced ultrafast rotation of the optical tensor
of a bulk WSe$_2$ crystal,
in a similar way as a Kerr rotation detected in monolayer TMDs
under magnetic fields.
The WSe$_2$ crystal is oriented as sketched in Fig.~\ref{fig:1}a and 
the probe incoming electric field is fixed oriented at $45$ degrees in the $x$-$y$ plane.
The polarization of the out-coming reflectivity
is hence detected simultaneously along the horizontal (H) zigzag direction,
and along the vertical (V) armchair direction.
Due to such a geometrical configuration,
the reflectivity along H- and V-directions
is identical under equilibrium conditions.
As a pump pulse, we use the laser fundamental at $\hbar\omega_{\rm pump}=1.55$ eV, which is nearly-resonant to the first A-exciton state.
The pump polarization can be quickly switched between left-circular (LCP)
and right-circular (RCP) by means of a liquid-crystal variable waveplate.
The incident fluence was set to 500 $\mu$J/cm$^2$. $\Delta R/R$ denotes the normalized differential reflectivity,
measured as a function of the pump-probe delay $t$ and the probe photon energy $\hbar\omega$. 

\begin{figure*}[t]
\includegraphics[width=16.5cm ]{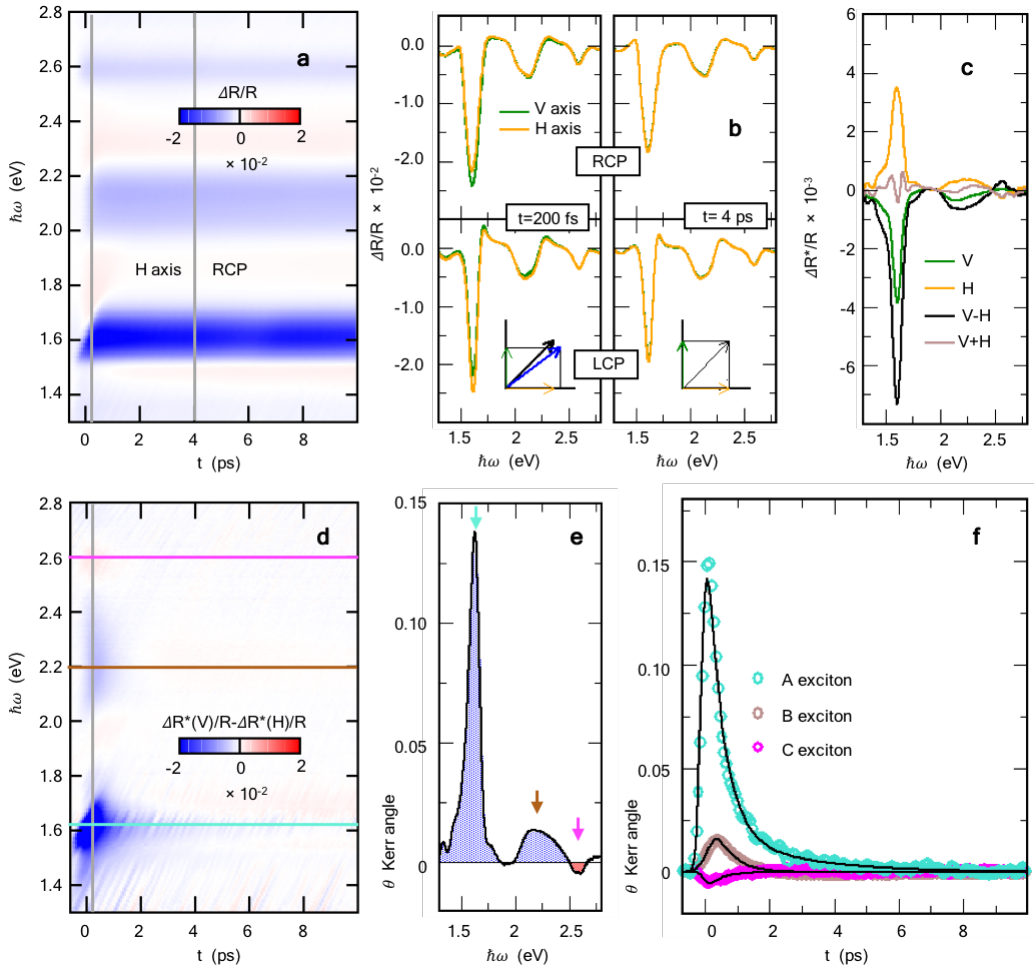}
\caption{
\textbf{Ultrafast pump-driven Kerr rotation in bulk WSe$_2$.}
\textbf{a},  Exemplary $\Delta R/R$ map for the H probe component, acquired with RCP pump polarization
(the other combinations are reported in Sec. ED1 of the  Extended Data),
showing typical
spectral features for the A, B and C excitons, at energies
$\hbar \omega \approx 1.6$, $2.2$, and $2.6$ eV.
The vertical grey lines at $t=200$ fs and $t=4$ ps mark the delays at which the spectral analysis,
shown in panel b, has been performed.
\textbf{b}, 
Spectral profiles of $\Delta R/R$ for the different combinations of pump and probe polarizations
at $t=200$ fs and $t=4$ ps. 
The insets shows a sketch of the corresponding rotation of the reflected beam polarization. 
\textbf{c}, Dichroism profiles $\Delta R^*/R$ extracted from panel (b) at $t=200$ fs
along the V and H directions.
Also shown are their sum and difference.
In particular, the difference signal $\Delta R^*/R(V)-\Delta R^*/R(H)$
is proportional to the optical Kerr rotation angle.
The negligible signal $\Delta R^*/R(V)+\Delta R^*/R(H)$ 
provides a check for the validity of the present analysis.
\textbf{d},
Energy vs. time color map of the optical dichroism $\Delta R^*/R(V)-\Delta R^*/R(H)$,
which is proportional to the rotation angle of the reflected polarization, triggered upon switching the helicity of the pump pulse polarization;
\textbf{e},
Net Kerr rotation angle $\theta$ as a function of the probe photon energy
$\hbar \omega$ at $t=200$ fs (vertical grey line in panel d);
\textbf{f},
Ultrafast dynamics of the Kerr angle $\theta$
at the three representative features (colored symbols corresponding to horizontal cuts in panel d).
Also shown are the corresponding fits (black solid lines).
}
\label{fig:2}
\end{figure*}

Fig.~\ref{fig:2}a shows one of the four $\Delta R/R$ maps acquired simultaneously, specifically for H probe polarization and RCP pump polarization
(the other maps being reported in Extended Data Fig. ED1).
Such spectra are sensitive to the broad-frequency effects of the total pump-driven photo-excitations in both valleys from the valence to the conduction bands.
Sharp features with negative $\Delta R/R$ at $\hbar \omega = \Delta_{\rm A} \approx 1.6$ eV and $\hbar \omega = \Delta_{\rm B} \approx 2.0-2.2$ eV signalize the A and B excitonic states, respectively \cite{Pogna_2016}.
Here we note an overall predominance of negative $\Delta R/R$, pointing out that such spectral features are governed by a true modification of the optical intensity rather than by an energy shift due to the screening modulation.
A clear feature is also detected at $\hbar \omega = \Delta_{\rm C} \approx  2.6$ eV, at the energy corresponding to the C exciton in bulk WSe$_2$ \cite{li14}.
The onset of these features has been rationalized in single-layer TMDs
in terms of the combined effect
of Pauli blocking and of pump-driven bandgap renormalization \cite{Pogna_2016}.
The time scale of the relaxation dynamics of all these features is
typically slow ($\sim 34-44$ ps, depending on which feature), consistent with previous literature \cite{shi13,cui14,zhu14,Mai_2014,Pogna_2016,yan17},
and it reflects the fact that the energy gap imposes a bottleneck to the recombination of electrons and holes.

In order to extract the appropriate Kerr signal, we analyze the normalized differential reflectivity $\Delta R/R$
for different circular polarization of the pump photon and for different H/V orientations.
The $\Delta R/R$ spectra for these four configurations are shown in Fig.~\ref{fig:2}b
for time delay $t=200$ fs and $t=4$ ps, corresponding to vertical grey lines in Fig.~\ref{fig:2}a.
At $t=4$ ps the normalized differential reflectivity $\Delta R/R$
does not display any sensitivity to the RCP/LCP pump polarization.
The observed optical profile of $\Delta R/R$ in this regime
can be thus attributed to bandgap renormalization induced
by pump-modified electronic screening.
We note that  $\Delta R/R$
has a sightly different profile along the H/V probe directions.
This fact is likely due to a small finite incidence angle ($\lesssim$ 4$^{\circ}$)
of the probe beam, that makes the two polarization states slightly inequivalent. 
For each given H or V direction,
a significant dependence on the circular pump polarization
is on the other hand reported at $t=200$ fs,
especially close to each of the three excitonic energies.
This is mostly evident close to the A-exciton resonance at $\hbar \omega=1.6$ eV, where in the V direction RCP gives rise to larger $| \Delta R/R |$ than LCP.
The polarization-induced changes in $\Delta R/R$
are reversed along the H direction,
with RCP leading in that case to a smaller $|\Delta R/R|$ at $\hbar \omega=1.6$ eV than LCP, with a reasonably good symmetry.
A similar picture is found to be valid 
in the energy ranges $\hbar \omega \approx 2.2$ eV and $\hbar \omega \approx 2.6$ eV, corresponding to the B and C excitons, although the magnitude of the effect in this case is strongly reduced.
Such reversible sensitivity on the circular polarization of the pump
cannot be attributed to a pure change of the electronic screening.
On the phenomenological ground, the present observations
can be recast
in terms of a modified optical conductivity tensor ${\bm \sigma}$,
with $2 \times 2$ components in the $x$-$y$ space \cite{levallois15}.
In the absence of pumping (or at large delay, e.g. $t=4$ ps), ${\bm \sigma}$
is purely diagonal with $\sigma_{xx}=\sigma_{yy}$.
The observed pump-driven changes in $\Delta R/R$
can be thus rationalized
in terms of a pump-driven onset of
a off-diagonal term with $\sigma_{xy}=-\sigma_{yx}\neq 0$,
giving rise to a net (Kerr) rotation of the linear probe polarization soon after the
photo-excitation,
as sketched in the insets of Fig.~\ref{fig:2}b.
The Kerr nature of the observed dichroism
is evident from Fig. \ref{fig:2}c 
where we plot the dichroism of $\Delta R/R$
[$\Delta R^*/R=\Delta R/R_{\rm RCP}-\Delta R/R_{\rm LCP}$]
at $t=200$ fs along the V and H directions.
We note that the energy-shape of $\Delta R^*/R$
along the H probe direction
is nicely matched by a corresponding
opposite $\Delta R^*/R$ along the V direction.
Such analysis is assessed at a quantitative level in the same figure
where we further show
the sum, $\Delta R^*/R(V)+\Delta R^*/R(H)$,
and the difference,
$\Delta R^*/R(V)-\Delta R^*/R(H)$,
between the two axes.
While the latter quantity can be directly interpreted as a pump-induced Kerr signal,
the first one represents what {\em it is not} Kerr effect.
Its negligibility with respect to the Kerr signal,
$|\Delta R^*/R(V)+\Delta R^*/R(H)| \ll |\Delta R^*/R(V)-\Delta R^*/R(H)|$,
provides thus a robust validation of our results.
Equipped with such spectral analysis for bulk WSe$_2$,
we plot in Fig. \ref{fig:2}d the energy-time dependence
of the optical dichroism $\Delta R^*/R(V)-\Delta R^*/R(H)$
which is proportional to the Kerr angle.
For comparison, we performed
a similar analysis
for bulk WS$_2$ and MoTe$_2$
at the same pump energy $\hbar\omega_{\rm pump} \approx 1.55$ eV
(which for these materials is not resonant with the A-exciton), and
no trace of any Kerr optical response
was found
in the whole energy range investigated
(see Extended Data Figs. ED2, ED3).

Three important features are worth being pointed out in the data for bulk WSe$_2$:
($i$) we show evidence of a sizable optical rotation at {\em all}
the three excitonic frequencies $\hbar\omega \approx 1.6$, $2.2$, $2.6$ eV,
well above the pumping energy $\hbar\omega_{\rm pump} \approx 1.55$ eV,
pointing out that pumping at the A-exciton resonance has a profound impact
on the entangled optical properties of all these features in a wide frequency range;
($ii$) we observe the same sign for the optical dichroism (i.e. Kerr rotation)
for both the A and B exciton spectra, but an {\em opposite} sign
for the C-exciton. 
This can be better visualized in Fig. \ref{fig:2}e where
we plot the Kerr angle $\theta(t,\hbar\omega)$ at $t=200$ fs 
(vertical line cut in Fig. \ref{fig:2}d)
as obtained by using the procedure outlined in Ref. \onlinecite{perlangeli20}
(Methods).
We characterize each Kerr feature with the energies
$\hbar\omega_i=1.63$, $2.2$ and $2.6$ eV where its maximum $|\theta^{\rm max}(t=200\,\mbox{fs},\omega)|$ occurs.
The ratio of the corresponding intensities at $\hbar\omega_i$ ($i=$A, B, C)
is found $|\theta_{\rm B}^{\rm max}|/|\theta_{\rm A}^{\rm max}| \approx 10 \%$,
$|\theta_{\rm C}^{\rm max}|/|\theta_{\rm A}^{\rm max}| \approx 3.4 \%$.
As we will see later, the relative sign among
these features can be rationalized and reproduced by means of a microscopical
quantum-field analysis, revealing the role of the valley/spin/layer quantum entanglement
and of the interlayer coupling;
($iii$) although the exciton population dynamics is found to decay
with a long time-scale $\tau_{\rm long} \sim 30-40$ ps (see Fig. \ref{fig:2}a),
the optical Kerr rotation at the relevant energies $\hbar\omega_i$ displays much shorter, but still sizable lifetimes.
The time evolution of $\theta(t,\hbar\omega_i)$
at the three characteristic energies  [horizontal cuts in panel (d)] is shown in Fig. \ref{fig:2}f.
In order to assess the characteristic timescales of the three features we fit each dataset with a standard fitting procedure, including the pump-probe cross-correlation driven risetime and one or two exponential decays
(see the Methods).
The fit curves are superimposed as black solid lines over the experimental points in Fig. \ref{fig:2}f.
We find that the time dynamics of the Kerr angle $\theta(t,\hbar\omega)$
at the A-exciton energy $\hbar \omega=1.63$ eV cannot be
described by a single exponential timescale,  whereas a bi-exponential decay
$\theta_{\rm A}(t) \propto a_1 \exp[-t/\tau_{\rm A,fast}]+a_2 \exp[-t/\tau_{\rm A, slow}]$ provides a good agreement, with a large component $a_1$ characterized by a fast dynamics $\tau_{\rm A,fast} \approx 476 \pm 30$ fs, followed by a weaker
component ($a_2/a_1 \approx 0.12$)
with a slower dynamics $\tau_{\rm A, slow} \approx 2315 \pm 50$ fs.
The analysis of the time dynamics at the B, C exciton energies
displays similar fast timescales with $\tau_{\rm B} \approx 528 \pm 30$ fs
and $\tau_{\rm C} \approx 511 \pm 30 $ fs, while no slower component
for these features is detected.

\section*{Discussion}

The optical Kerr rotation upon circularly polarized pumping
implies a time-reversal-symmetry breaking and it
has been theoretically predicted and observed in monolayer TMDs 
\cite{Wang_2013,Mai_2014,plechinger14,zhu14,dalconte15,yang15,yan17,mccormick18,kiemle20,catarina20,arora21}
where it has been attributed
to the lack of a center of symmetry, together with
the valley-selective population induced by the
circularly polarized pumping.
To our knowledge, the present observation is the first direct
report of a net Kerr effect in {\em bulk} TMDs
with crystal  inversion symmetry.  
The natural questions are then: how
a Kerr rotation is activated by circularly-polarized pumping
in bulk TMDs without breaking the point-inversion-symmetry?
Why similar spectral fingerprints
can be found in both (non-centrosymmetric) monolayer and (centrosymmetric) bulk TMDs?

To address these issues, we performed a quantum-field analysis (Methods)
of the optical response in the presence of
circularly polarized photo-induced particle-hole excitations
resonant with the A-exciton energy.
In order to capture all the observed
Kerr spectral features,
we consider a non-interacting model, that is enough to capture
the relevant physics, and
we employ a six-band ${\bf k}\cdot {\bf p}$  model
that generalizes
to the bulk case 
the three-band 
model,
previously employed for single-layer TMDs \cite{3bands,rcc}.
Within this context only relevant $d$-orbitals
of the metal atoms, with atomic orbital angular
momentum $l_{\rm at}=0,\pm 2$,
are retained,
providing a good modelling of the relevant
valence and
conduction bands,
as well as of a third block of bands
with $l_{\rm at}=2$ $d$-orbitals,
associated with the high-energy excitons
and which have been proposed
to be responsible for the C-exciton peak.
Following previous literature,
we assume a local and spin-diagonal
spin-orbit coupling \cite{3bands,roldan14},
and
we expand the total Hamiltonian
close to the high-symmetry points $\nu$K ($\nu=\pm$) of the bulk Brillouin zone,
so that
$\hat{H}({\bf k})\approx \sum_{s,\nu} \hat{H}_{s,\nu}({\bf p})$,
where $\hat{H}_{s,\nu}$ 
is defined in the $6 \times 6$
orbital space of bulk systems,
$s=\pm1$ is the spin,
$\nu=\pm 1$ represents the valley index $\pm$K, 
and ${\bf p}={\bf k}-\nu {\rm K}$ is the relative momentum
with respect to the corresponding valley.

Before discussing the full spectral
properties of the Kerr response, it
is worth clarify the underlying mechanisms giving rise to a finite Kerr rotation.
To this purpose
we focus for the moment on
the static limit of the off-diagonal optical tensor
$\sigma_{xy}(\hbar \omega \to 0)$, where the Kerr response
extrapolates smoothly to the Hall rotation.
Within this framework, it is known that the term $\sigma_{xy}(0)$
can be conveniently related to the underlying topological properties
encoded in the Berry curvature $\Omega_{n,s}({\bf k})$, i.e.
$ \sigma_{xy} = (e^2/\hbar)
\sum_{n,s,\nu}
\sum_{\bf p}
\Omega_{n,s,\nu}({\bf p})
f[E_{n,s,\nu}({\bf p})]$,
where $n$ is the band index
and $f[x]$ the momentum/band population factor
that, at equilibrium, corresponds to the thermal Fermi-Dirac distribution.
We neglect for the moment the
high-energy conduction bands,
which play a role only for high-energy
spectral features, and
we consider the reduced
$4 \times 4$ Hamiltonian
where only the low-energy conduction
and valence bands are retained,
with an effective interlayer hopping term which mainly couples the valence bands
(Methods).
Such Hamiltonian has been 
previously conveniently used
to underline the intrinsic entanglement between the valley, the layer and
the spin polarization \cite{gong13},
which can conveniently
be defined in the orbital/layer space.
This approach however
cannot be generalized in a straightforward way
to the Hall/Kerr response
since the Berry curvature
is defined in the band-space.

\begin{figure*}[t]
\includegraphics [width=1\textwidth ]{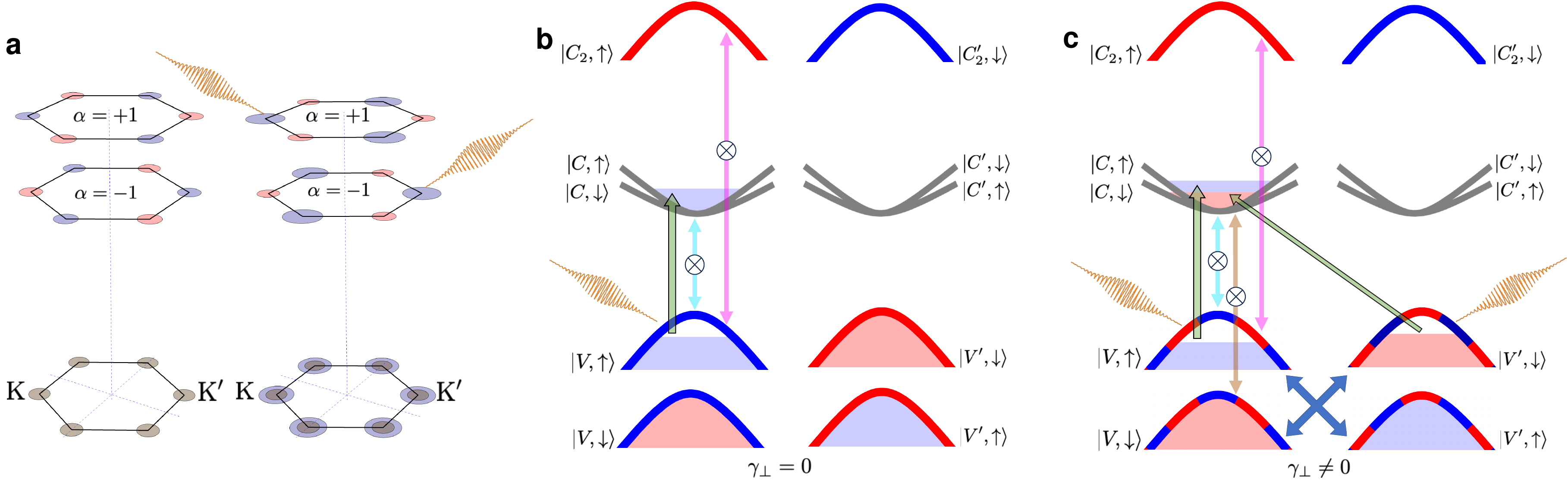}
\caption{\textbf{Hidden order and effects of selection rules on the Kerr optical response.}
\textbf{a}, Schematic representation of the different contributions to the $\sigma_{xy,\alpha,\nu}$ Kerr optical rotation resolved for layer and valleys.
For sake of representation, the spin degree has been averaged.
The bottom line represent the total layer-averaged $\sigma_{xy,\nu}=\sum_\alpha \sigma_{xy,\alpha,\nu}$.
Blue/red areas represent states resulting in a positive/negative Kerr angle, respectively, while grey areas stand for null Kerr rotation.
Left panel shows the hidden order before pumping.
LCP laser pumping in the right panel photo-excites charges at K in layer $\alpha=+1$,
and at -K in layer $\alpha=-1$.
In both cases it results in a net
increase of $\sigma_{xy}>0$ in each valley, resulting
in a finite Kerr rotation. The system still obeys inversion symmetry
and each valley is equivalent.
\textbf{b}, Representative selection rules
and Pauli-blocked optical transitions at $\nu=1$ valley
in the absence of interlayer coupling.
The absorption of a LCP photon tuned at the A-energy
triggers particle-hole transitions (upward green arrow)
only from the 
$E_{{\rm V},\uparrow}$ state with  $l_{\rm at}=2$ (blue bands)
to  $E_{{\rm C},\uparrow}$ with  $l_{\rm at}=0$ (grey bands).
The resulting photo-excited charges $n_{{\rm V},\uparrow}=n_{{\rm C},\uparrow}\neq 0$
induce Pauli blocking (crossed colored vertical double-arrows)
at both the A- and C-resonance energies.
\textbf{c},
Same as panel b in
the presence of a finite interlayer coupling,
which
leads to a mixed $l_{\rm at}$ character for the
$E_{{\rm V},s}$ bands, resulting
in a large charge transfer from $E_{{\rm V},\uparrow}$ to  $E_{{\rm C},\uparrow}$,
but also in a  finite charge transfer from
$E_{{\rm V},\downarrow}$ to  $E_{{\rm C},\downarrow}$,
hence $n_{{\rm V},\downarrow}=n_{{\rm C},\downarrow}\neq 0$.
A Pauli-blocking is consequently active also for
the particle-hole transitions responsible for the B-resonance.
}
\label{fig:3}
\end{figure*}

Despite this apparent limitation, in the following
we show that in bulk TMDs
it is possible to define
in a compelling way
a layer-projection
of the Berry curvature
for each band.
As detailed in Methods section, this task is
made feasible in bulk TMDs due to the layered electronic structure {\em and} to
the specific form of
the interlayer coupling which is negligible for the conduction bands.
Within this context,
using the above
Hamiltonian 
$\hat{H}_{s,\nu}({\bf p})$,
we can thus evaluate
a layer/spin/valley-resolved
Berry curvature 
$\Omega_{n,s,\nu,\alpha}({\bf p})$,
which allows us to compute
as well a layer/spin/valley-resolved off-diagonal response
$\sigma_{xy,s,\nu,\alpha}$,
with $\alpha=\pm1$ standing for the layer index (Methods). 
Under equilibrium conditions,
we obtain at the valley points ${\bf p}=0$:
\begin{equation}
\sigma_{xy,s,\nu,\alpha}^{\rm eq}
=-\frac{e^2}{\hbar}\left[
\tilde{\lambda}
\left(
\Omega_{\rm A}-\Omega_{\rm B}
\right)
s
+
\left(
\Omega_{\rm A}+\Omega_{\rm B}
\right)
\nu \alpha\right],
\label{sigma0}
\end{equation}
where $\Omega_{\rm A}$
is the Berry curvature
of the top valence band
at the valley point
(associated with the A-exciton optical transitions),
$\Omega_{\rm B}$
is the Berry curvature
of the bottom valence band
(associated with the B-exciton optical transitions),
and $\tilde{\lambda}=\lambda/\sqrt{\lambda^2+\gamma_\perp^2}$ is a dimensionless parameter expressing the relative relevance of the
spin-orbit coupling $\lambda$ with respect to the
interlayer coupling $\gamma_\perp$.
Eq. (\ref{sigma0})
displays the layer/spin/valley-resolved hidden-order of the Kerr/Hall response.
The first term in Eq. (\ref{sigma0}) descends directly
from the analysis of
the 
finite spin-resolved
Berry curvature,
without the need of a layer-projection,
and it reflects the possible onset
of a finite Hall/Kerr rotation
upon breaking of the time-reversal
symmetry \cite{catarina20}.
Obeying the space-inversion symmetry, it does not depend
on the valley index $\nu$.
The second term in Eq. (\ref{sigma0}) is on the other hand a novel feature which
appears only within
a layer-resolved analysis
and
it reveals that in bulk TMDs
at equilibrium,
although a net 
(layer-averaged) valley-Hall (valley-Kerr) response
is null,
a hidden order is present,
with a finite valley/layer 
Hall conductivity response $\sigma_{xy,\nu,\alpha}^{\rm eq} \neq 0$
as a consequence of the underlying entanglement
between valley $\nu$
and layer $\alpha$.
A representative plot
of the hidden order of
$\sigma_{xy,\nu,\alpha}^{\rm eq}$
is shown in Fig.  \ref{fig:3}a,
showing a staggered pattern
in the layer/valley space.
Such layer/valley entanglement
plays a crucial role
in our context since
it points out an alternative
path for generating a finite Hall/Kerr response in bulk TMDs, not related to a direct time-reversal-symmetry breaking, but as a consequence
of a hidden {\em inversion-symmetry breaking}.

To understand how a circularly-polarized pumping
can exploit such hidden order in inducing a finite Kerr rotation, we compute 
at the valley points (${\bf p}=0$) the optical selection rules,
which can be related to the Berry curvature through the valuable expression \cite{Xiao_rmp_2010}:
\begin{align}
    {\cal P}^{\zeta}_{n\to m,s,\nu}
    &=
 \left|J^{nm}_{x,s,\nu} (0)\right|^2/2 + \left|J^{nm}_{y,s,\nu} (0)\right|^2/4
\nonumber\\
&
 - \zeta
    \left[ E_{n,s,\nu} (0)-E_{m,s,\nu} (0)\right]^2 \Omega_{nm,s,\nu}(0)/4
    .
    \label{ourselec}
\end{align} 
Eq. (\ref{ourselec})
is proportional to the probability
of absorbing a photon with chiral polarization $\zeta=\pm$
accompanied by
a particle-hole excitation from the $n$-band to the $m$-band
with spin $s$ in the valley $\nu$.
On the ground of the above discussion, and as detailed
in Methods section, we recognize
that the layer indices do not
mix in Eq. (\ref{ourselec})
so that we can properly define
{\em layer-resolved}
selection rules
${\cal P}^{\zeta}_{n\to m,s,\nu,\alpha}$ (Methods).
Joining ${\cal P}^{\zeta}_{n\to m,s,\nu,\alpha}$
with the detailed expressions
of the layer/spin/valley-resolved
Berry curvature 
$\Omega_{n,s,\nu,\alpha}$, we can finally derive
the total effect
on the off-diagonal optical
response of the
absorption
of a circularly-polarized
photon tuned at the A-exciton edge:
\begin{eqnarray}
\delta\bar{\sigma}_{xy,s,\nu,\alpha}^\zeta
&=&
\delta\sigma_{xy}^{\rm dichro}
\zeta
+
\delta\sigma_{xy}^{\rm full}
\zeta s \nu \alpha
\nonumber\\
&&
\delta\sigma_{xy}^{\rm orb}
\nu \alpha
+
\delta\sigma_{xy}^{\rm spin}
s
.
\label{dichrofullm}
\end{eqnarray}
The first term in (\ref{dichrofullm})
is the only one that survives
after averaging over
all the internal degrees of freedom,
and it is the one
responsible for the Kerr response
observed in our experimental setup.
It arises as consequence of two different channels: 
the effect of 
the selection rules of a chiral photon on the spin sector, $\propto \zeta s$,
which acts on top of the
spin structure of the hidden order
$\propto s$ [first term in Eq. (\ref{sigma0})];
and
the effect of the selection rules induced by a chiral photon on the orbital order, $\propto \zeta \nu \alpha$,
exploiting the underlying hidden order $\propto \nu \alpha$
[second term in Eq. (\ref{sigma0})].
A schematic picture of the onset of a finite Hall/Kerr
response from such entanglement is shown in 
the right panel of Fig. \ref{fig:3}a:
a circularly-polarized photon
will act at the $\nu=+1$ valley
mainly on layer $\alpha=+1$ and in similar way
at $\nu=-1$ 
on layer $\alpha=-1$.
The net result is an off-diagonal term
with a well define sign $\delta\sigma_{xy}>0$
in both valleys, obeying thus the
lattice inversion symmetry.
The third and forth terms in
(\ref{dichrofullm}) on the other hand describe the
effects of the pumping that do not
give rise to any dichroism,
as related for instance 
to
the change of the overall charge distribution.
They obey the same symmetries
of the bulk system and they
follow thus the same pattern of the hidden order.
They can be probed in the presence
of an explicit space/time breaking
by external fields.
Finally, the second term of (\ref{dichrofullm}) conveys
the full entanglement among all the degrees of freedom.
The above scenario is consistent with previous ARPES investigations of the hidden
valley/helicity/spin entanglement in bulk TMDs \cite{riley14,bertoni16,razzoli17,beaulieu20,dong21,fanciulli23}.

After having clarified the mechanisms giving
rise to a light-induced Kerr response in centrosymmetric bulk
TMDs,
we investigate now the fundamental information encoded
in the spectral profile of the optical Kerr rotation,
with a particular focus on the origin of the three main features
outlined in Fig. \ref{fig:2}e and on their relative sign.
We evaluate the optical conductivity
$\sigma_{ij}(\omega)$ ($i,j=x,y$)
within a non-interacting Kubo formalism
using the full
$6 \times 6$ Hamiltonian, retaining thus
also the high-energy conduction bands.
As shown in Methods,
three main contributions can
be identified,
$\sigma_{ij}^{\rm A}(\omega)$,
$\sigma_{ij}^{\rm B}(\omega)$,
$\sigma_{ij}^{\rm C}(\omega)$,
associated with the
interband transitions responsible for the
A-, B- and C-exciton optical features,
respectively.
The off-diagonal part of
each of these terms fulfills at a spectral level,
under equilibrium conditions,
the same symmetries as their static limit
in Eq. (\ref{sigma0}),
and it vanishes in the absence
of symmetry breaking or of external fields.

Finite spectral features $\delta\sigma_{xy}(\omega)$
appear however upon the effect of circularly-polarized pumping
tuned at the A-exciton resonance.
For standard values of fluence, pump-driven particle-hole excitations are localized
very close to valleys ${\bf p}=0$,
modifying in a non-thermal way the population
factors $f[E_n]\approx f^{\rm eq}[E_n]+\delta f[E_n]$.
The selection rules reported
in Eq. (\ref{ourselec}) provide
a useful guidance for evaluating such effects.
More in details, we obtain:
\begin{eqnarray}
    {\cal P}^{\zeta}_{{\rm V}\to {\rm C},s,\nu} (0) 
    &=& 
(1-\tilde{\lambda}\zeta s-\nu \zeta+\tilde{\lambda}s\nu)/4
,
\label{mysel1m}
\\
    {\cal P}^{\zeta}_{{\rm V}\to {\rm C}^\prime,s,\nu} (0) 
    &=& 
(1-\tilde{\lambda}\zeta s+\nu \zeta-\tilde{\lambda}s\nu)/4
,
\label{mysel2m}
\end{eqnarray} 
where ${\cal P}^{\zeta}_{{\rm V}\to {\rm C}}$,
${\cal P}^{\zeta}_{{\rm V}\to {\rm C}^\prime}$
describe the probability
of transition from the top valence band V
to the (degenerate)
low-energy conduction bands C, C$^\prime$.
Using such information within
the Kubo computation of the off-diagonal
optical response,
we can estimate the integrated spectral area
$I_{\rm K}^i=
   \sum_{s,\nu}
   \int   {\rm Im} \,\,
\delta \sigma_{xy,s,\nu}^{i}(\omega)   d\omega$,
for each of the optical Kerr features:
\begin{eqnarray}
I^{\rm A}_{\rm K}
&\propto&
-
\frac{v_1^2}{\Delta_{\rm A}}
(1+3\tilde{\lambda}^2) \zeta ,
\label{intam}
\\
I^{\rm B}_{\rm K}
&\propto&
-
\frac{v_1^2}{\Delta_{\rm B}}
(1-\tilde{\lambda}^2) \zeta ,
\label{intbm}
\\
I^{\rm C}_{\rm K}
&\propto&
\frac{v_3^2}{\Delta_{\rm C}}
\tilde{\lambda}(1+\tilde{\lambda}) \zeta 
,
\label{intcm}
\end{eqnarray}
where the factors $v_1$, $v_3$
are related to the current matrix element
between different bands.

Noticeable information are encoded 
in Eqs. (\ref{intam})-(\ref{intcm}):
($i$) they predict that the Kerr rotation
at the A- and B-resonances have the same sign,
and an opposite sign with respect to the Kerr rotation at the C-edge, in perfect
agreement with the experimental observations;
($ii$) the onset of a Kerr response at the B-resonance in bulk TMDs
under circularly-polarized pumping
is intimately related to the interlayer coupling, and it would vanish
in the decoupled-layer limit $\tilde{\lambda}\to 1$;
($iii$) on the other hand, the spin-orbit
coupling has a fundamental role in
the Kerr feature at the C-edge, 
and it would disappear
in the opposite limit of no spin-orbit
(or extremely large interlayer coupling)
$\tilde{\lambda}\to 0$.
On the quantitative level, taking
from literature the values
of $\tilde{\lambda}=0.959$,\cite{gong13}
and of the ratio $v_3/v_1=0.733$,\cite{rcc}
we estimate $|I^{\rm A}_{\rm K}|
> |I^{\rm B}_{\rm K}| > |I^{\rm C}_{\rm K}|$,
with
$|I^{\rm B}_{\rm K}|/|I^{\rm A}_{\rm K}|
\approx 0.24$
and
$|I^{\rm C}_{\rm K}|/|I^{\rm A}_{\rm K}|
\approx 0.16$.
Note that such estimates are based on a non-interacting analysis and do not take
into account thus
many-body renormalization and other effects).
More precise first-principle analyses,
using GW-BSE calculations,
predict for instance
a much smaller ratio of the optical oscillation strengths $v_3^2/v_1^2\approx 0.04$ \cite{lin21}, in even further agreement
with the experimental observation.

We can achieve a
more microscopical insight
on the processes responsible for each spectral feature,
and on their time-dynamics,
by relating, using the Kubo formalism,
the Kerr spectral intensity at different
energies to pump-driven photo-excited charge densities $n_{n,s}$.
In particular, focusing at a single valley $\nu=1$ (in bulk systems
both valley are equivalent)
and for left-circularly-polarized photons, we get:
\begin{align}
I^{\rm A}_{\rm K}(t)
\propto &
\frac{v_1^2}{\Delta_{\rm A}}
\left[
c^2_{\lambda}
(
n_{{\rm C},\uparrow}
-
n_{{\rm C}^\prime,\downarrow}
)
-
s^2_{\lambda}
(
n_{{\rm C}^\prime,\uparrow}
-
n_{{\rm C},\downarrow}
)
\right.
\nonumber\\
&
\left.
+
(c^2_{\lambda}-s^2_{\lambda})
(
n_{{\rm V},\uparrow}
-
n_{{\rm V},\downarrow}
)
\right]
,
\label{iam}
\\
I^{\rm B}_{\rm K}(t)
\propto &
\frac{v_1^2}{\Delta_{\rm B}}
\left[
s^2_{\lambda}
(
n_{{\rm C},\uparrow}
-
n_{{\rm C}^\prime,\downarrow}
)
+
c^2_{\lambda}
(
n_{{\rm C},\downarrow}
-
n_{{\rm C}^\prime,\uparrow}
)
\right]
,
\label{ibm}
\\
I^{\rm C}_{\rm K}(t)
\propto &
-
\frac{v_3^2}{\Delta_{\rm C}}
c^2_{\lambda}
(
n_{{\rm V},\uparrow}
-
n_{{\rm V},\downarrow}
)
.
\label{icm}
\end{align}
Here the prefactors
$c^2_{\lambda}=(1+\tilde{\lambda})/2$ and
$s^2_{\lambda}=(1-\tilde{\lambda})/2$
take into account the 
(time-independent)
magnitude of the optical activities
for given interband transitions,
whereas the population $n_{n,s}(t)$ can have a time dynamics.

Let us first consider as a template case
the limit of decoupled layers $\gamma_\perp=0$
in the bulk 2H structure.
In that case
$c^2_{\lambda}=1$,
$s^2_{\lambda}=0$.
Keeping in mind that the two layers of the units
cell are rotated each other by 180$^\circ$,
the electronic structure of the bulk systems
is thus obtained by the degenerate superposition
of the electronic states $\psi_n({\bf k},\alpha=+1)$  and of
the electronic states
$\psi_n({\bf k},\alpha=-1)$.
In this limit
the bulk bands are thus
completely layer-localized,
with bands of layer $\alpha=+1$
corresponding to the K point
of the single-layer system,
and bands of layer $\alpha=-1$
corresponding to K$^\prime$.
At $t=0$
the absorption of a LCP photon
with energy tuned at
the A-exciton edge $\hbar \omega \approx \Delta_{\rm A}$
triggers at $\nu=1$
particle-hole transitions only in layer $\alpha=+1$
(Fig. \ref{fig:3}b).
Considering also the role of the geometrical phase winding of the Bloch states
$l_{\rm geo}$ \cite{Xiao2008,cao12,Rostami2015},
particle-hole transitions are allowed only from
the valence state $E_{{\rm V},\uparrow}$,
with atomic orbital  angular momentum $l_{\rm at}=2$ (blue bands)
to the conduction state
$E_{{\rm C},\uparrow}$
with atomic orbital angular momentum
$l_{\rm at}=0$ (grey bands) \cite{cao12},
with a net total angular momentum change
$\Delta l_{\rm tot}=-1$.
The only finite photo-excited charge densities
will be thus completely spin-polarized
$n_{{\rm C},\uparrow}=
n_{{\rm V},\uparrow} \neq 0$.
On the other hand,
no particle-hole transitions
are allowed from
the valence state $E_{{\rm V},\downarrow}$
to the conduction band
$E_{{\rm C}^\prime,\downarrow}$,
although fulfilling the energy-resonant requirement, because
$E_{{\rm V},\downarrow}$ is localized
on the layer $\alpha=-1$ and it borrows
from these in-layer states
the main orbital angular momentum $l_{\rm at}=-2$ (red bands) which is not
coupled with the absorption of a LCP photon.
The concomitant Pauli blocking of the
conduction 
and valence bands, 
$E_{{\rm C},\uparrow}$,
$E_{{\rm V},\uparrow}$
with a change of
the selected orbital angular momentum
$\Delta l _{\rm tot}=-1$,
gives rise to the finite Kerr response
at the A-edge energy,
with an overall spectral area as
in Eq. (\ref{iam}).
On the other hand,
the photo-excited hole density
in the valence
band $E_{{\rm V},\uparrow}$
is responsible also for
the Pauli blocking of
the optical transitions
between the valence
band $E_{{\rm V},\uparrow}$
and the high-energy
conduction band
$E_{{\rm C}_2,\uparrow}$,
associated with the C-exciton shoulder.
However,
due to the different orbital character
of conduction bands C$_2$ ($l_{{\rm at}}=2$)
with respect to
conduction bands C ($l_{{\rm at}}=0$),
such Pauli blocking involves
a change of total angular momentum
$\Delta l _{\rm tot}=1$ {\em opposite}
to the one involved in the A-resonance,
and hence a Kerr effect at
$\hbar \omega=\Delta_{\rm C}$
with {\em opposite sign}
than at $\hbar \omega=\Delta_{\rm A}$.
The very experimental observation
of a Kerr response at the C-edge,
furthermore with the correct sign
as predicted by the theory,
supports thus the identification of the broad C-exciton feature as induced by
a strong band nesting for optical transitions between the valence
and the high-energy conduction bands at the
valley points.

It is worth noticing here that,
within the assumption of
no interlayer coupling,
no Kerr response at the B-resonance
is expected as direct effect
of circularly-polarized pumping.
A  finite Kerr rotation at the B-exciton edge upon circularly-polarized pumping
has been reported in single-layer TMDs, where it was 
explained as an effect of the
valley depolarization induced by intervalley scattering \cite{Mai_2014,yang15,yan17}.
We rule out this possible explanation
of our experimental observation
since
it will imply an {\em opposite} sign of the Kerr angle at $\hbar\omega \approx \Delta_{\rm B}$
with respect to $\hbar\omega \approx \Delta_{\rm A}$.
Our theoretical analysis [Eq. (\ref{intbm})]
shows however
that a finite Kerr response at the
B-resonance, with the correct sign
as experimentally observed,
is a direct product the circularly-polarized
pumping, even
in the absence of many-body scattering processes.
The physics describing this effect
is mathematically captured in Eq. (\ref{ibm}) and it is schematically
summarized in Fig. \ref{fig:3}c.
The fundamental feature to be highlighted here is that
the interlayer coupling
hybridizes states in different layers with
opposite orbital angular momentum. 
The mixed orbital-angular-momentum character
is depicted in Fig. \ref{fig:3}c
as dashed blue-red bands.
The remarkable aspect is that at the valley $\nu=+1$, the
photo-induced particle-hole
excitations, and the corresponding photo-charges,
can occur not only
between $E_{{\rm V},\uparrow}$
and $E_{{\rm C},\uparrow}$ (thick vertical green arrow),
but also, with a minor extent,
between $E_{{\rm V},\downarrow}$
and $E_{{\rm C},\downarrow}$ (tiny oblique green arrow).
The corresponding photo-induced
charges at $t=0$ can be estimated as
$n_{{\rm C},\uparrow}=c^2_{\lambda}$,
$n_{{\rm C},\downarrow}=s^2_{\lambda}$, 
$n_{{\rm V},\uparrow}=c^2_{\lambda}$, $n_{{\rm V},\downarrow}=s^2_{\lambda}$, 
whereas
$n_{{\rm C}^\prime,\uparrow}=n_{{\rm C}^\prime,\downarrow}=0$.
The Kerr response
at the B-resonance is thus induced
by the two parallel channels:
the Pauli blocking
of the
$E_{{\rm V}^\prime,\uparrow} \leftrightarrow
E_{{\rm C},\uparrow}$
transitions,
driven by the large conduction
charge $n_{{\rm C},\uparrow}$,
but with a small current matrix element
$s^2_{\lambda}$
induced by the interlayer coupling;
and the Pauli blocking
of the
$E_{{\rm V}^\prime,\downarrow} \leftrightarrow
E_{{\rm C},\downarrow}$
transitions,
driven by the small conduction
charge $n_{{\rm C},\downarrow}$,
but with a large (mainly intralayer) current matrix element $s^2_{\lambda}$.
Both these channels trigger
an angular momentum change
$\Delta l _{\rm tot}=-1$,
just at the processes
associated with the A-resonance,
and induce thus a Kerr response
at the B-edge with the same
sign as at the A-resonance.

Equipped with Eqs. (\ref{iam})-(\ref{icm}),
having identified the origin and the physical processes
for all the three spectral features in the Kerr response, we comment on
the observed time-dependence.
In order to determine the quantum entanglement
betweent the different degrees of freedom and the optical
selection rules, Eqs. (\ref{iam})-(\ref{icm})
[and (\ref{mysel1m})-(\ref{mysel2m})]
have been derived using a non-interacting
model.
Such features are not expected to
change where the many-body interactions,
and the actual formation of bound excitons
are taken into account, with the obvious
warning that exciton resonance energies
should be considered as inferred
from the experiments, including thus
the many-body band renormalization and the
exciton binding energy.
Eqs. (\ref{iam})-(\ref{icm})
provide thus a useful guidance
to rationalize the charge time-dynamics,
including the effective time-dependence
$n_{n,s}(t)$.
Starting from the pump-induced
populations at $t=0$,
$n_{{\rm C},\uparrow}(0)=
n_{{\rm V},\uparrow}(0)=
c^2_{\lambda}$,
$n_{{\rm C},\downarrow}=
n_{{\rm V},\downarrow}=s^2_{\lambda}$, 
$n_{{\rm C}^\prime,\uparrow}(0)=
n_{{\rm C}^\prime,\downarrow}(0)=0$,
the overall Kerr response fades out
either when spin-flip/layer-flip
processes lead to equal spin/layer
populations at the valley points
(e.g.
$n_{{\rm V},\uparrow}\approx
n_{{\rm V},\downarrow}$,
$n_{{\rm C},\uparrow}
\approx n_{{\rm C}^\prime,\uparrow}$);
or when, in bulk TMDs, photo-excited charges
migrate to other points of the
Brillouin zone, forming for instance
an excitons.
Eqs. (\ref{iam})-(\ref{icm})
appear thus very powerful since
the analysis
of the time-dependence of the Kerr response
at the A-, B- and C-edges allows
in principle to trace in an independent way
the time-dynamics of the
conduction ($I_{\rm K}^{\rm B}$)
and valence ($I_{\rm K}^{\rm C}$) charges,
as well as their joint effect
($I_{\rm K}^{\rm A}$).
On this ground, it is strikingly
remarkable that all the three Kerr features
display a similar time-decay
with $\tau_{\rm fast} \approx 500$ fs
(leaving aside the weak long-lived
component of the A feature).
This observation
strongly suggests thus
that our set-up,
with energy pumping tuned at the A-resonance,
does {\em not} induce free itinerant charges
$n_{{\rm V},s}$,
$n_{{\rm C},s}$,
$n_{{\rm C}^\prime,s}$
in an independent way
in the conduction and valence bands,
but it simply affects the total
number of available excitons.
Photo-induced
conduction and valence charges
are thus still bound together,
with the time-scale $t \lesssim 500$ fs, 
by the exciton binding energy,
and they obey thus the same
time dynamics, hampering also
the independent migration towards different
points of the Brillouin zone.
The common decay time
that we observe for all the three
Kerr features 
at room temperature 
is comparable with the reported time-dynamics of the Kerr A-feature in monolayer WSe$_2$ at room temperature \cite{zhu14,molina17,yan17}, 
suggesting thus that the observed
time-dynamics reveals the time decay due
to the
spin-flip/layer-flip
processes, which are similar as in monolayer
systems (note that, due to the 2H stacking,
layer-flip processes in bulk TMDs
are similar to inter-valley
processes in monolayer compounds).
Time-resolved angle-resolved photoemission spectroscopy (tr-ARPES) explores
on the other hand
a different physics, tracing the
dynamics of the
(unbound) charges left
by the photoemission process
(in the valence bands or in the pump-populated conduction bands).
Not bound in the exciton pairs,
these charges can optimize
their energy
by easily migrating to their own minima
in bulk samples (Q-point for conduction,
$\Gamma$-point for valence bands),
with a much faster dynamics \cite{bertoni16,molina17,dong21,fanciulli23,rcc}.
Our results present thus an additional
information and they
are thus
not incompatible with tr-ARPES data
in bulk WSe$_2$,
which report a much shorter
time decay ($\tau_{\rm B, ARPES} \approx 100$ fs) for
the dichroism in the conduction bands  \cite{bertoni16,dong21,fanciulli23}.

In summary, we have reported evidence of a sizable Kerr optical rotation,
driven by circularly polarized light, in a wide energy range on bulk WSe$_2$.
The present results show that efficient optical manipulation
is affordable not only in single-layer TMDs with broken inversion symmetry,
but also in centrosymmetric bulk WSe$_2$,
circumventing the limitation of atomically-thin samples.
We rationalize the presence of a pump-induced Kerr effect in centrosymmetry bulk TMDs as a result of the time-reversal-symmetry breaking
induced by the circularly-polarized light
absorption exploiting the
hidden entanglement between the orbital content, valleys and layers.
This scenario
allows for a finite Kerr rotation
in bulk samples without a selective valley population. 
We also argue that the joint analysis
of the spectral intensities of the
three Kerr features provides a powerful
insight on the dynamics
of the photo-excited particle-hole
excitations.
On this ground our findings
suggest that bound excitons
can live for a rather long
time-scale ($\tau \approx 500$ fs)
also in bulk TMDs.
Similar dichroism and Kerr effects could be exploited in the bulk structure
of other transition metal dichalcogenides such as Mo-based compounds. 
Furthermore, the known presence of a large spin-orbit coupling
in WSe$_2$ implies that a sizable spin polarization is also possible
in bulk TMDs,
in agreement thus with the findings of Ref. \onlinecite{fanciulli23}.

\bibliography{manuscript.bib}

\section*{Methods}

{\small

\subsection*{Samples and Experiments}

High quality bulk single-crystals of 2H-WSe$_2$ were purchased from HQ Graphene. Their in-plane orientation was determined by ex-situ LEED (Low Energy Electron Diffraction) experiments, and the samples were mounted in the optical setup accordingly to this information. TR-OS experiments were performed using the experimental setup described in \cite{perlangeli20}. Briefly, the output of a Ti:Sapphire regenerative Amplifier (Coherent RegA), delivering ~50 fs pulses at 800 nm (1.55 eV) and 250 kHz, is splitted for obtaining pump and probe pulses. A broadband supercontinuum pulse extending in a range 400-1100 nm and linearly-polarized at 45 degrees with respect to the optical table is obtained by focusing ~1 $\mu$J/pulse of energy ($\approx$250 mW of power) in a 3 mm thick Sapphire window. The reflected beam from the sample is separated in its horizontal (H) and vertical (V) polarization components and spectrally dispersed by a Wollaston Polarizer, and finally detected by a pair of 512-pixels CMOS linear array detectors by Hamamatsu. The pump pulse at 800 nm is polarized circularly with a liquid crystal waveplate from Thorlabs, that allows to quickly ($\approx$50 ms) switch between the two circular polarization states. The spot size on the sample amounts to 160 um and 50 um FWHM for the pump and the probe respectively. The two beams cross at an angle of 20 degrees.

\subsection*{Evaluation of Kerr angle from reflectivity anisotropy}

We describe here the procedure employed to estimate the Kerr rotation angle
from the measurement of the reflectivity anisotropy upon circularly polarized pumping.

For this task, we essentially followed the scheme nicely described in Refs. \onlinecite{levallois15,levallois12}.

Using the quantum field theory , we first computed microscopically,
as described in the previous Section, the optical conductivity tensor
\begin{equation}
\hat{\sigma}(\omega)
=
\left(
\begin{matrix}
\sigma_{xx}(\omega) & \sigma_{xy}(\omega) \\
-\sigma_{xy}(\omega) & \sigma_{xx}(\omega)
\end{matrix}
\right),
\end{equation}
where we have assumed $\sigma_{yy}(\omega)=\sigma_{xx}(\omega)$,
and hence the complex dielectric function
$\hat{\epsilon}(\omega)$
through the standard relation
$\hat{\epsilon}(\omega)=\epsilon_\infty \hat{I}+4\pi i \sigma_{xx}(\omega)/\omega$.

In our setup, we shed incoming light at 45 degrees in the $x$-$y$ plane, thus with
the same components $E_x=E_y$.
The optical response along the along $x$ (H) and $y$ (V)
is governed thus by:
\begin{eqnarray}
\epsilon_{\rm V}(\omega)
&=&
\epsilon_{xx}(\omega) + \epsilon_{xy}(\omega)
,
\\
\epsilon_{\rm H}(\omega)
&=&
\epsilon_{xx}(\omega) - \epsilon_{xy}(\omega)
.
\end{eqnarray}
The corresponding reflectivity along the two axes was thus determined as:
\begin{equation}
r_{\alpha}(\omega)
=
\frac{1-\sqrt{\epsilon_{\alpha}(\omega)}}{1+\sqrt{\epsilon_{\alpha}(\omega)}},
\end{equation}
and the absolute value of the reflectivity as:
\begin{equation}
R_\alpha(\omega)
=
\left|
r_{\alpha}(\omega)
\right|^2
,
\end{equation}
where $\alpha=$H, V.

The Kerr angle $\theta(\omega)$
has been eventually computed through the standard formula \cite{perlangeli20}:
\begin{equation}
\theta(\omega)
=
\frac{1}{2}
\frac{R_{\rm H}(\omega)-R_{\rm V}(\omega)}
{R_{\rm H}(\omega)+R_{\rm V}(\omega)}
.
\end{equation}

\subsection*{Time-dependence fitting procedure}

The time-traces have been fitted using the conventional approach, making use of one or two exponential decays of the form $a_i \exp(-t/t_i)$, $i=1,2$, where $a_i$ is the amplitude of the exponential component, $t_i$ its the corresponding time-constant and $t$ represents the pump-probe delay. The data have been fitted using Igor Pro and performing the convolution of the above expression with a Gaussian function accounting for the finite temporal resolution of the setup, due to the pump-probe cross correlation. The Gaussian FWHM used is 300 fs, that is mostly determined by the large pump-probe crossing angle.

\subsection*{Quantum field Theory analysis 
}

For a detailed description of the relevant features
of bulk TMDs,
we generalized for bulk TMDs
the three-band ${\bf k}\cdot {\bf p}$ model,
previously employed for single-layer TMDs \cite{3bands,rcc}.
Within this context only the relevant $d$-orbitals of the metal atoms with atomic orbital angular
momentum $l_{\rm at}=0,\pm 2$
are retained,
providing a good modeling of the low-energy
conduction bands $E_0({\bf p})$ with dominant
$l=0$ character;
of the valence bands $E_{\rm B}({\bf p})$,
which correspond actually at the bottom block
of bands with dominant $|l_{\rm at}|=2$ character;
and of a block of high-energy conduction bands,
$E_{\rm T}({\bf p})$,
with dominant $|l_{\rm at}|=2$ character, which are responsible
for higher energy optical
transitions \cite{lin21,lin22,rcc}.
Taking into account that
the two $MX_2$ layers of the bulk unit cell
have a relative rotation of 180 degrees,
a suitable Hilbert space
at the valley $\nu$ was provided by
$\psi_{s,\nu}^\dagger=(\psi_{s,\nu,1}^\dagger,\psi_{s,\nu,-1}^\dagger)$ where
\begin{align}
\psi_{s,\nu,\alpha}^\dagger
=
(
d_{0,s,\alpha}^\dagger,
d_{{\rm B},s,\nu,\alpha}^\dagger,
d_{{\rm T},s,\nu,\alpha}^\dagger
)    
.
\end{align}
Here $d_{0,s,\alpha}^\dagger$
creates in layer $\alpha$ an electron with spin $s$ in the
orbital $d_{3z^2-r^2}$;
$d_{{\rm B},s,\nu,\alpha}^\dagger$
creates in layer $\alpha$ an electron with spin $s$ 
and with chiral orbital linear combination $d_{x^2-y^2}+ i \nu d_{xy}$;
$d_{{\rm T},s,\nu,\alpha}^\dagger$
creates in layer $\alpha$ an electron with spin $s$ 
and with chiral orbital linear combination $d_{x^2-y^2}-i \nu d_{xy}$.
In this basis 
the bulk Hamiltonian reads:
\begin{eqnarray}
    \hat{H}_{s,\nu}({\bf p})
    &=&
    \begin{pmatrix}
    \hat{h}_{s,\nu}({\bf p}) & \hat{h}_\perp \\
    \hat{h}_\perp^\dagger & \hat{h}_{s,-\nu}({\bf p})
     \end{pmatrix} 
    ,
    \label{Hbulkp}
\end{eqnarray}
where ${\bf p}={\bf k}-\nu K$
is the relative momentum with respect
to the valley,
and $\hat{h}_{s,\nu}({\bf p})$
is the ${\bf k}\cdot {\bf p}$ Hamiltonian for the single-layer \cite{rcc}:
\begin{align}
 \hat{h}_{s,\nu}({\bf p})
    &=
    \begin{pmatrix}
         E_0(p) & \hbar v_1 p_{\nu,-} & \hbar v_2 p_{\nu,+}
     \\
     \hbar v_1 p_{\nu,+}  & E_{\rm B}(p)+\lambda s\nu & \hbar v_3 p_{\nu,-}
     \\ 
    \hbar v_2 p_{\nu,-}   & \hbar v_3 p_{\nu,+}  & E_{\rm T}(p)-\lambda s\nu
     \end{pmatrix} 
    ,
\end{align}
with $p_{\nu,\pm}=\nu p_x\pm ip_y$
and $\lambda$ the spin-orbit coupling.
The only (spin-diagonal)
interlayer term in $\hat{h}_\perp$
is known to be the coupling $\gamma_\perp$
between the
block of $E_{\rm B}$ bands.
Other interlayer couplings mix $E_0$ and
$E_{\rm T}$ with the large energy
difference \cite{cappelluti13},
and their effect can be neglected.
The properties of the bulk
system were conveniently
investigated in the rotated basis that diagonalizes
Eq. (\ref{Hbulkp}) at ${\bf p}=0$,
\begin{align} \label{npsib2}
\phi_{s,\nu}^\dagger
&=
(
d_{{\rm C},s,\nu}^\dagger,
d_{{\rm V},s,\nu}^\dagger
d_{{\rm C}_2,s,\nu}^\dagger,
d_{{\rm C}^\prime, s,\nu}^\dagger,
d_{{\rm V}^\prime , s,\nu}^\dagger,
d_{{\rm C}_2^\prime, s,\nu}^\dagger,
)
,
\end{align}
where 
\begin{eqnarray}
d_{{\rm C},s,\nu}
&=&
d_{0,s,1},
\label{tri}
\\
d_{{\rm C}^\prime , s,\nu}
&=&
d_{0,s,-1},
\\
d_{{\rm V},s,\nu}
&=&
\cos\chi_{s,\nu}
d_{{\rm B},s,\nu,1}
+
\sin\chi_{s,\nu}
d_{{\rm B},s,-\nu,-1},
\\
d_{{\rm V}^\prime, s,\nu}
&=&
-\sin\chi_{s,\nu}
d_{{\rm B},s,\nu,1}
+
\cos\chi_{s,\nu} 
d_{{\rm B},s,-\nu,-1}
,
\\
d_{{\rm C}_2,s,\nu}
&=&
d_{{\rm T},s,\nu,1}
,
\\
d_{{\rm C}_2^\prime, s,\nu}
&=&
d_{{\rm T},s,-\nu,-1}
.
\label{trf}
\end{eqnarray}
Here $\chi_{s,\nu}=(1/2)\arccos(\tilde{\lambda} \nu s)$,
where we defined $\tilde{\lambda}=\lambda/\sqrt{\lambda^2+\gamma_{\perp}^2}$.
For sake of compactness, we define also
$c_{s,\nu}=\cos\chi_{s,\nu}$ and $s_{s,\nu}=\sin\chi_{s,\nu}$.
In the basis (\ref{npsib2}) the Hamiltonian reads:
\begin{eqnarray}
    \hat{\cal H}_{s,\nu}({\bf p}) 
    &=&
    \begin{pmatrix}
    \hat{\cal h}_{s,\nu}({\bf p}) & \hat{\cal h}_\perp \\
    \hat{\cal h}_\perp^\dagger & \hat{\cal h}^\prime_{s,\nu}({\bf p})
     \end{pmatrix} 
    ,
    \label{Hbulkp2}
\end{eqnarray}
where
\begin{eqnarray}
 \hat{\cal h}_{s,\nu}({\bf p})
    &=&
    \begin{pmatrix}
         E_{{\rm C},s,\nu}(p) & \hbar v_1 c_{s,\nu} p_{\nu,-} & \hbar v_2 p_{\nu,+}
     \\
    \hbar v_1 c_{s,\nu} p_{\nu,+}  & E_{{\rm V},s,\nu}(p) & \hbar v_3 c_{s,\nu} p_{\nu,-}
     \\ 
    \hbar v_2 p_{\nu,-}  &  \hbar v_3 c_{s,\nu} p_{\nu,+} & E_{{\rm C}_2,s,\nu}(p)
     \end{pmatrix} 
    ,
\end{eqnarray}
\begin{eqnarray}
 \hat{\cal h}^\prime_{s,\nu}({\bf p})
    &=&
    \begin{pmatrix}
         E_{{\rm C}^\prime,s,\nu}(p) & \hbar v_1 c_{s,\nu} p_{-\nu,-} & \hbar v_2 p_{-\nu,+}
     \\
    \hbar v_1 c_{s,\nu} p_{-\nu,+}  & E_{{\rm V}^\prime,s,\nu}(p) & \hbar v_3 c_{s,\nu} p_{-\nu,-}
     \\ 
    \hbar v_2 p_{-\nu,-}  &  \hbar v_3 c_{s,\nu} p_{-\nu,+} & E_{{\rm C}_2^\prime,s,\nu}(p)
     \end{pmatrix} 
    ,
\end{eqnarray}
and
\begin{eqnarray}
 \hat{\cal h}_{\cal \perp}
    &=&
    \begin{pmatrix}
         0 & \hbar v_1 s_{s,\nu} p_{-\nu,+} & 0
     \\
    \hbar v_1 s_{s,\nu} p_{-\nu,+}  & 0 & \hbar v_3 s_{s,\nu} p_{-\nu,-}
     \\ 
    0  &  \hbar v_3 s_{s,\nu} p_{-\nu,-} & 0
     \end{pmatrix} 
    .
    \label{lasth}
\end{eqnarray}
The current operators 
$\hat{J}_{i,s,\nu}({\bf p})=
\partial \hat{\cal H}_{s,\nu}({\bf p})/\partial p_i$,
can be also promptly derived
in the band basis
from Eq. (\ref{Hbulkp2}). 

The Hamiltonian in Eq. (\ref{Hbulkp2})
defines at ${\bf p}=0$ 
the two split valence bands,
$E_{{\rm V},s,\nu}=
E_{\rm B}+\sqrt{\lambda^2+\gamma_\perp^2}$,
$E_{{\rm V}^\prime,s,\nu}(0)=
E_{\rm B}-\sqrt{\lambda^2+\gamma_\perp^2}$,
with mixed-layer character.
They correspond
to the antibonding and bonding interlayer states,
respectively.
On the other hand,
due to the negligible interlayer coupling,
the states of the low-energy
and high-energy conduction bands,
are strongly localized on each layer,
in accordance with Eqs. (\ref{tri})-(\ref{trf}),
with 
$E_{{\rm C},s,\nu}=E_{{\rm C}^\prime,s,\nu}
=E_0$ being essentially degenerate,
while $E_{{\rm C}_2,s,\nu}$
$E_{{\rm C}^\prime,s,\nu}$
display a sizable spin-orbit splitting,
$E_{{\rm C}_2,s,\nu}=E_{\rm T}-\lambda s\nu$,
$E_{{\rm C}_2^\prime,s,\nu}=E_{\rm T}+\lambda s\nu$.
The energy differences:
$\Delta_{\rm A}=E_0-E_{\rm V}$,
$\Delta_{\rm B}=E_0-E_{\rm V^\prime}=$,
$\Delta_{\rm C}=E_{\rm T}-\lambda-E_{\rm V}$,
define the edges, in the non-interacting limit, for the
particle-hole continuum associated with the A- B- and C-exciton resonances, respectively.

{\em Layer-resolved  selection rules
and layer/spin/valley/photon entanglement in bulk TMDs}
-
In order to analyze
the topological character and
the optical selection rules of bulk TMDs
we focused on the $4 \times 4$
reduced relevant Hilbert space
where only the valence and low-energy conduction bands were retained
\cite{gong13,liureview15,korma15},
neglecting at this stage the additional high-energy conduction bands.

The Berry curvature for each band was computed as
$\Omega_{n,s,\nu} ({\bf p}) = \sum_{m \neq n}    \Omega_{nm,s,\nu} ({\bf p})$,
where\cite{liureview15}
\begin{equation}
    \Omega_{nm,s,\nu} ({\bf p}) 
    =
    -
    2{\rm Im}
    \left[
    \frac{
    \langle  J_{x,s,\nu} ({\bf p}) \rangle_{n,m} \langle  J_{y,s,\nu} ({\bf p}) \rangle_{m,n}
    }
    {[E_{n,s,\nu}({\bf p})-E_{m,s,\nu}({\bf p})]^2}
    \right].
\label{berryband}
\end{equation}
Here $\langle  J_i \rangle_{n,m}$
is the matrix element of the
expectation value of
the current operators $\hat{J}_i$
expressed in the band-basis (\ref{npsib2}),
and we have implicitly used the fact that
the Hamiltonian (\ref{Hbulkp2}) is diagonal
in both the spin and valley degrees of freedom.

For small pump fluences, 
the light-induced particle-hole excitations are
localized nearby the valley points ${\bf p}=0$,
so that the physical properties are well captured
by an analysis at ${\bf p}=0$.
We got in particular:
\begin{eqnarray}
\Omega_{{\rm C},s,\nu}(0)
&=&
\left(
\Omega_{\rm A}+\Omega_{\rm B}
\right)
\nu
+
\tilde{\lambda}
\left(
\Omega_{\rm A}-\Omega_{\rm B}
\right)
s,
\label{oca}
\\
\Omega_{{\rm C}^\prime,s,\nu}(0)
&=&
-
\left(
\Omega_{\rm A}+\Omega_{\rm B}
\right)
\nu
+
\tilde{\lambda}
\left(
\Omega_{\rm A}-\Omega_{\rm B}
\right)
s
,
\\
\Omega_{{\rm V},s,\nu}(0)
&=&
-2\tilde{\lambda}\Omega_{\rm A} s
,
\\
\Omega_{{\rm V}^\prime,s,\nu}(0)
&=&
2\tilde{\lambda}\Omega_{\rm B} s
.
\label{ovpa}
\end{eqnarray}
where
$\Omega_{\rm A}=\hbar^2v^2/\Delta_{\rm A}^2$
is the  Berry curvature associated with the
particle-hole transitions responsible for the A-exciton edge,
and $\Omega_{\rm B}=\hbar^2v^2/\Delta_{\rm B}^2$
is the  Berry curvature associated with the
particle-hole transitions responsible for the B-exciton edge.

Within the same framework, we derive the
optical selection rules
governing the absorption of a photon with chiral polarization $\zeta=\pm$
driving a particle-hole excitation from the $n$-band to the $m$-band
\cite{Xiao_rmp_2010}:
\begin{align}
    {\cal P}^{\zeta}_{n\to m,s,\nu}
    &=
 \left|J^{nm}_{x,s,\nu} (0)\right|^2/2 + \left|J^{nm}_{y,s,\nu} (0)\right|^2/4
\nonumber\\
&
 - \zeta 
    \left[ E_{n,s,\nu} (0)-E_{m,s,\nu} (0)\right]^2 \Omega_{nm,s,\nu}(0)/4
    .
\end{align} 
In the specific case of a pump energy tuned
at the A-exciton edge, i.e.
accompanied by a particle-hole excitation from 
the valence band V to the conduction band C or C$^\prime$, we get
the dimensionless expressions:
\begin{eqnarray}
    {\cal P}^{\zeta}_{{\rm V}\to {\rm C},s,\nu} (0) 
    &=& 
(1-\tilde{\lambda}\zeta s-\nu \zeta+\tilde{\lambda}s\nu)/4
,
\label{mysel1}
\\
    {\cal P}^{\zeta}_{{\rm V}\to {\rm C}^\prime,s,\nu} (0) 
    &=& 
(1-\tilde{\lambda}\zeta s+\nu \zeta-\tilde{\lambda}s\nu)/4
.
\label{mysel2}
\end{eqnarray} 

Further insight was gained by splitting
the current operators in their layer-projected
components,
$\hat{J}_{i,s,\nu}=\sum_\alpha \hat{J}_{i,s,\nu,\alpha}$.
It should be remarked that,
in general, this projection
is {\em not} sufficient for defining
{\em layer-projected} Berry curvatures and
{\em layer-projected} selection rules,
since different layer components can mix
in Eq. (\ref{berryband}).
A careful analysis showed however that at ${\bf p}=0$
such mixing does not occur in our case
thanks to the layered structure
of bulk TMDs and to the reduced
effect of the interlayer coupling.
We could thus introduce
in a compelling way a {\em layer-resolved}
Berry curvature $\Omega_{n,s,\nu,\alpha}(0)$
such that
$\Omega_{n,s,\nu}(0) =\sum_\alpha \Omega_{n,s,\nu,\alpha}(0)$.
The possibility of defining a layer-resolved Berry curvature
allowed us to treat in a compelling way the layer index
as an additional quantum degree of freedom
and to evaluate
{\em layer-resolved} (as well a spin/valley-resolved)
 topological properties, selection rules
and, in an ultimate analysis,
a {\em layer-resolved}
off-diagonal optical response $\sigma_{xy,s, \nu,\alpha}$
through the relation: 
\begin{align}
\sigma_{xy,s,\nu,\alpha}
&=
-
\frac{e^2}{\hbar}
\sum_{{\bf p},n}
\Omega_{n,s,\nu,\alpha}({\bf p})
f[E_{n,s,\nu}({\bf p})]
,
\label{kerrst}
\end{align}
where $f[E]$ is the population factor (non necessarily thermal).
In order to investigate the
entanglement between the different degrees
of freedom, since the Berry curvatures
are commonly peaked at the valley points,
we focus in detail on the contribution at ${\bf p}=0$,
where
$\sigma_{xy,s,\nu,\alpha}
\approx
-
(e^2/\hbar)
\sum_{{\bf p},n}
\Omega_{n,s,\nu,\alpha}(0)
f[E_{n,s,\nu}(0)]$.
Using the layer-projected current operators,
we obtain:
\begin{align}
\Omega_{{\rm C},s,\nu,\alpha}(0)
&=
\left[
\left(
\Omega_{\rm A}+\Omega_{\rm B}
\right)
\nu
+
\tilde{\lambda}
\left(
\Omega_{\rm A}-\Omega_{\rm B}
\right)
s
\right]
\frac{1+\alpha}{2}
,
\label{ocl}
\\
\Omega_{{\rm C}^\prime,s,\nu,\alpha}(0)
&=
\left[
-\left(
\Omega_{\rm A}+\Omega_{\rm B}
\right)
\nu
+
\tilde{\lambda}
\left(
\Omega_{\rm A}-\Omega_{\rm B}
\right)
s
\right]
\frac{1-\alpha}{2}
,
\\
\Omega_{{\rm V},s,\nu,\alpha}(0)
&=
-\Omega_{\rm A} \nu \alpha
-\tilde{\lambda}\Omega_{\rm A} s
,
\\
\Omega_{{\rm V}^\prime,s,\nu,\alpha}(0)
&=
-
\Omega_{\rm B} \nu \alpha
+
\tilde{\lambda}\Omega_{\rm B} s
.
\label{ovpl}
\end{align}
Under equilibrium conditions we obtain thus: 
\begin{eqnarray}
\sigma_{xy,s,\nu,\alpha}^{\rm eq}
&=&
-
\left(
\Omega_{\rm A}+\Omega_{\rm B}
\right)
\nu \alpha
-
\tilde{\lambda}
\left(
\Omega_{\rm A}-\Omega_{\rm B}
\right)
s
.
\end{eqnarray}
The pattern of $\sigma_{xy,s,\nu,\alpha}^{\rm eq}$ is depicted
in Fig. 3a and it displays the intrinsic hidden order of the Kerr response
including the spin, valley and layer degrees of freedom.
For a bulk-sensitive spin-integrated probe,
as our setup in the absence of pumping,
we get $\sigma_{xy}^{\rm eq}=\sum_{s,\nu,\alpha} \sigma_{xy,s,\nu,\alpha}^{\rm eq}=0$.

The layer-resolved Berry curvature allowed us
to derive also the following layer-resolved
optical selection rules:
\begin{eqnarray}
    {\cal P}^{\zeta}_{{\rm V}\to {\rm C},s,\nu,\alpha} (0) 
    &=& 
   (1-\tilde{\lambda}\zeta s-\nu \zeta+\tilde{\lambda}s\nu) 
    (1+\alpha)/8
,
\label{pvcl}
\\
   {\cal P}^{\zeta}_{{\rm V}\to {\rm C}^\prime,s,\nu,\alpha} (0) 
    &=& 
(1-\tilde{\lambda}\zeta s+\nu \zeta-\tilde{\lambda}s\nu)
 (1-\alpha)/8
.
\label{pvcpl}
\end{eqnarray} 
On the ground of Eqs. (\ref{pvcl})-(\ref{pvcpl}),
along with Eqs. (\ref{ocl})-(\ref{ovpl}),
we could furthermore evaluate the additional contribution
$\delta \bar{\sigma}_{xy}^\zeta$
to the off-diagonal response induced
by the absorption of circularly-polarized pumping.
Retaining the explicit dependence of all the degrees
of freedom, after a careful analysis we get:
\begin{eqnarray}
\delta\sigma_{xy,s,\nu,\alpha}^\zeta
&\propto&
-
\frac{2\Omega_{\rm A}+\Omega_{\rm B}
+
\tilde{\lambda}^2
(2\Omega_{\rm A}-\Omega_{\rm B})
}{4}
\zeta
\nonumber\\
&&
 +
 \frac{2\Omega_{\rm A}+\Omega_{\rm B}
+
\tilde{\lambda}
(2\Omega_{\rm A}-\Omega_{\rm B})
}{4}
\nu \alpha
\nonumber\\
&&
+
\frac{\tilde{\lambda}\Omega_{\rm A}}{2} s
-
\tilde{\lambda}
\Omega_{\rm A}
\zeta \alpha s\nu
.
\label{dichrofull}
\end{eqnarray}

{\em Optical Kerr response}
-
In order to achieve a detailed understanding of all observed optical Kerr features,
it was important to retain all the optically-active bands,
as captured by the Hamiltonian (\ref{Hbulkp2}).
In similar way as for single-layer TMDs,
we split  thus the total
response in three interband
contributions \cite{rcc},
$\sigma_{xy}=\sigma_{xy}^{\rm v-c}
+\sigma_{xy}^{\rm v-c_2}+\sigma_{xy}^{\rm c-c_2}$.
The first term $\sigma_{xy}^{\rm v-c}$
accounts for interband transitions between
the block of valence bands (v$=$V, V$^\prime$)
and the block of low-energy conduction bands
(c$=$C, C$^\prime$),
and it is responsible for the spectral features
at the A-edge and B-edge exciton energies;
the second term $\sigma_{xy}^{\rm v-c_2}$
described the
interband transitions between
the block of valence bands (v$=$V, V$^\prime$)
and the block of high-energy conduction bands
(c$_2=$C$_2$, C$_2^\prime$).
Finally, the third term
describes optical transitions
between the block of low-energy conduction bands
and the block of high-energy conduction bands.
This term is Pauli-blocked in semiconducting bulk TMDs
at equilibrium. Although these optical features can be activated
upon pumping, they don't play a relevant role
in the present context and we neglected them.

The term $\sigma_{xy}^{\rm v-c}$
can be further divided 
in a contribution arising from the V bands,
as resulting in optical features at
the A-exciton edge,
$\sigma_{xy}^{\rm A}=\sigma_{xy}^{\rm V-c}$,
and
in a contribution arising from the V$^\prime$ bands,
as resulting in optical features at
the B-exciton edge,
$\sigma_{xy}^{\rm B}=\sigma_{xy}^{\rm V^\prime-c}$.
In similar way,
one can isolate in $\sigma_{xy}^{\rm v-c_2}$
the contribution 
$\sigma_{xy}^{\rm C}=\sigma_{xy}^{\rm V-c_2}$
associated with transitions
between the V bands and the band
C$_2$, C$_2^\prime$ with lowest energy
(this latter label depends
on the valley/spin index).
Using Eqs. (\ref{Hbulkp2})-(\ref{lasth})
and the Kubo formalism,
we can thus write:
\begin{eqnarray}
\sigma_{xy,s,\nu}^{\rm A}(\omega)
&=&
-
i \nu
\frac{e^2 v_1^2}{4\pi^2 \hbar^2 \omega}
\Big[
c^2_{s,\nu}
M(E_{{\rm C},s,\nu},E_{{\rm V},s,\nu},\omega)
\nonumber\\
&&
-
s^2_{s,\nu}
M(E_{{\rm C}^\prime,s,\nu},E_{{\rm V},s,\nu},\omega)
\Big]
,
\\
\sigma_{xy,s,\nu}^{\rm B}(\omega)
&=&
-
i \nu
\frac{e^2 v_1^2}{4\pi^2 \hbar^2 \omega}
\Big[
s^2_{s,\nu}
M(E_{{\rm C},s,\nu},E_{{\rm V}^\prime,s,\nu},\omega)
\nonumber\\
&&
-
c^2_{s,\nu}
M[E_{{\rm C}^\prime,s,\nu},E_{{\rm V}^\prime,s,\nu},\omega)
\Big]
,
\\
\sigma_{xy,s,\nu}^{\rm C}(\omega)
&=&
i \nu
\frac{e^2 v_3^2}{4\pi^2 \hbar^2 \omega}
\Big[
I_{s\nu}
c^2_{s,\nu}
M[E_{{\rm C}_2,s,\nu},E_{{\rm V},s,\nu},\omega)
\nonumber\\
&&
-
I_{-s\nu}
s^2_{s,\nu}
M[E_{{\rm C}_2^\prime,s,\nu},E_{{\rm V},s,\nu},\omega)
\Big]
,
\end{eqnarray}
where
\begin{eqnarray}
M(E_n,E_m,\omega )
&=&
\sum_{\bf p}
\left\{
\frac{f[E_n({\bf p})]-f[E_m({\bf p})]}{E_n({\bf p})-E_m({\bf p})-\hbar\omega -i\delta}
\right.
\nonumber\\
&&
\left.
-
\frac{f[E_n({\bf p})]-f[E_m({\bf p})]}{E_n({\bf p})-E_m({\bf p})+\hbar\omega+i\delta}
\right\}
,
\end{eqnarray}
and where
$I_{s\nu}=(1+s\nu)/2$ traces down
that the level $E_{\rm T}-\lambda$ is associated
with different bands C$_2$, C$_2^\prime$ for different valley/spin indices.
In semiconducting bulk TMDs at equilibrium,
$f[E_{{\rm V},s,\nu}({\bf p})]=1$, $f[E_{{\rm C},s,\nu}({\bf p})]=0$, 
$f[E_{{\rm C}_2,s,\nu}({\bf p})]=0$, and all the three features vanish
once summed over the spin.

Finite spectral features $\delta\sigma_{xy}(\omega)$
appear however upon the effect of circularly-polarized pumping
tuned at the A-exciton resonance.
For standard values of fluence, pump-driven particle-hole excitation are localized
very close to valleys ${\bf p}=0$,
modifying in a non-thermal way the population
factors $f[E]\approx f^{\rm eq}[E]+\delta f[E]$.
For energies $\hbar \omega$ close to the resonance edge,
we could evaluate the change in the response functions:
\begin{align}
\mbox{Im}\delta M(E_1,E_2,\omega)
&\approx
\pi
\delta[E_1(0)-E_2(0)-\hbar\omega]
\nonumber\\
&
\times
\sum_{\bf p}
\left\{
\delta f[E_1({\bf p})]-\delta f[E_2({\bf p})]
\right\}
\nonumber\\
&=
\pi
\delta[E_1(0)-E_2(0)-\hbar\omega]
\left\{\delta n_1 -\delta n_2
\right\}
,
\label{charges}
\end{align}
where $\delta n_i$ is the pump-driven charge variation
in the corresponding band $E_i$.
Using the selection rules Eqs. (\ref{mysel1})-(\ref{mysel2})
we can estimate
$\delta n_{{\rm V},s,\nu}=-(1-\tilde{\lambda}\zeta s)F/2$,
$\delta n_{{\rm C},s,\nu}=(1-\tilde{\lambda}\zeta s-\nu \zeta+\tilde{\lambda}s\nu)F/4$,
$\delta n_{{\rm C}^\prime,s,\nu}=(1-\tilde{\lambda}\zeta s+\nu \zeta-\tilde{\lambda}s\nu)F/4$,
where $F$ is a factor which scales linearly with the pump fluence.
For all the other bands, not affected by a pumping at the A-resonance, $\delta n=0$.
With such modeling we could evaluate thus the pump-driven 
off-diagonal part the optical tensor, $\delta \sigma_{xy,s,\nu}(\omega)$,
and the spectral integrated area
of each optical Kerr feature $i=$A, B, C as 
\begin{align}
   I_{\rm K}^i=
   \sum_{s,\nu}
   \int   {\rm Im} \,\,
\delta \sigma_{xy,s,\nu}^{i}(\omega)   d\omega
.
\end{align}

We get:
\begin{eqnarray}
I^{\rm A}_{\rm K}
&\propto&
-
\frac{v_1^2}{\Delta_{\rm A}}
(1+3\tilde{\lambda}^2) \zeta F,
\label{inta}
\\
I^{\rm B}_{\rm K}
&\propto&
-
\frac{v_1^2}{\Delta_{\rm B}}
(1-\tilde{\lambda}^2) \zeta F,
\label{intb}
\\
I^{\rm C}_{\rm K}
&\propto&
\frac{v_3^2}{\Delta_{\rm C}}
\tilde{\lambda}(1+\tilde{\lambda}) \zeta F
.
\label{intc}
\end{eqnarray}

Using the layer-resolved analysis,
we computed also the different spin/layer contributions to Eqs. (\ref{inta})-(\ref{intc}) .
Focusing on $\nu=1$, and considering for instance a left-circularly-polarized photon $\zeta=-1$,
we get:
\begin{eqnarray}
I_{s,\alpha=1}^{\rm A}
&\propto&
 \frac{v_1^2}{\Delta_{\rm A}} 
\left(
\frac{1+\tilde{\lambda}^2}{2}+\tilde{\lambda}s\right)F
,
\\
I_{s,\alpha=-1}^{\rm A}
&\propto&
-\frac{v_1^2}{\Delta_{\rm A}} 
\left(
\frac{1-\tilde{\lambda}^2}{4}
\right)F
,
\\
I_{s,\alpha=1}^{\rm B}
&\propto&
 \frac{v_1^2}{\Delta_{\rm B}} 
\frac{1-\tilde{\lambda}^2}{4}F
,
\\
I_{s,\alpha=-1}^{\rm B}
&\propto&
0,
\\
I_{s,\alpha=1}^{\rm C}
&\propto&
 -\frac{v_3^2}{\Delta_{\rm C}} 
\frac{(1+\tilde{\lambda})^2(1+s)}{8} F
,
\\
I_{s,\alpha=-1}^{\rm C}
&\propto&
\frac{v_3^2}{\Delta_{\rm C}} 
\frac{(1-\tilde{\lambda}^2)(1-s)}{8} F
.
\end{eqnarray}

Few interesting properties can be noticed here: ($i$) the Kerr spectral features
at the C-resonance $\hbar \omega = \Delta_{\rm C}$ arises from
a unique spin sector, and they are expected to be fully polarized.
On the contrary, the Kerr feature at the B-edge $\hbar \omega = \Delta_{\rm B}$
is totally spin-independent;
($ii$) the Kerr features at the B-edge
stem also from a unique layer;
($iii$) in the regime of weak interlayer coupling (which is realized in many bulk TMDs) the Kerr features at the A-resonance
$\hbar \omega = \Delta_{\rm A}$,
as well as the C-one,
are fully spin and layer polarized.
Eq. (\ref{charges}) allowed us to relate
the Kerr spectral intensity at different
energies to pump-driven photo-excited charge densities.
In particular, focusing at a single valley $\nu=1$, and for left-circularly-polarized photons, we get:
\begin{align}
I^{\rm A}_{\rm K}
\propto &
\,
\frac{v_1^2}{\Delta_{\rm A}}
\left[
c^2_{\lambda}
(
n_{{\rm C},\uparrow}
-
n_{{\rm C}^\prime,\downarrow}
)
-
s^2_{\lambda}
(
n_{{\rm C}^\prime,\uparrow}
-
n_{{\rm C},\downarrow}
)
\right.
\nonumber\\
&
\left.
+
(c^2_{\lambda}-s^2_{\lambda})
(
n_{{\rm V},\uparrow}
-
n_{{\rm V},\downarrow}
)
\right]
,
\\
I^{\rm B}_{\rm K}
\propto &
\,
\frac{v_1^2}{\Delta_{\rm B}}
\left[
s^2_{\lambda}
(
n_{{\rm C},\uparrow}
-
n_{{\rm C}^\prime,\downarrow}
)
+
c^2_{\lambda}
(
n_{{\rm C},\downarrow}
-
n_{{\rm C}^\prime,\uparrow}
)
\right]
,
\\
I^{\rm C}_{\rm K}
\propto &
\,
-
\frac{v_3^2}{\Delta_{\rm C}}
c^2_{\lambda}
(
n_{{\rm V},\uparrow}
-
n_{{\rm V},\downarrow}
)
,
\end{align}
where
$c^2_\lambda=(1+\tilde{\lambda})/2$ and
$s^2_\lambda=(1-\tilde{\lambda})/2$.

\vspace{5mm}

\section*{Acknowledgements}

E.C. acknowledges financial support from PNRR MUR Project No. PE0000023-NQSTI.

\mbox{}

\section*{Author contributions statement}

F.C. conceived and conducted the experiments.
E.C. and H.R. developed the theory.
All the authors conceived the paper
and contributed equally to the writing of the manuscript.

\section*{Competing interests}

The authors declare no competing interests.

\mbox{}

\section*{Additional information}

\textbf{Correspondence and requests for materials}
can be equally addressed to any of the Authors.

}

\end{document}

\sloppy

\newcommand{\blue}[1]{{\textcolor{blue} {#1}}}
\newcommand{\red}[1]{{\textcolor{red} {#1}}}
\newcommand{\green}[1]{{\textcolor{green} {#1}}}
%Title of paper

\title{Extended Data for: \\ Ultrafast light-driven optical rotation and hidden orders in bulk WSe$_2$}

\author{Emmanuele Cappelluti}
\email{emmanuele.cappelluti@ism.cnr.it}
\affiliation{Istituto di Struttura della Materia, CNR (CNR-ISM), 34149 Trieste, Italy}

\author{Habib Rostami}
\email{hr745@bath.ac.uk}
\affiliation{Department of Physics, University of Bath, Claverton Down, Bath BA2 7AY, United Kingdom}

\author{Federico Cilento}
\email{federico.cilento@elettra.eu}
\affiliation{Elettra-Sincrotrone Trieste S.C.p.A., 34149 Basovizza, Italy}

%\date{\today}

\maketitle
%\onecolumngrid
%\tableofcontents 
%\clearpage
%\newpage

\begin{figure}
\includegraphics[width=13.5cm ]{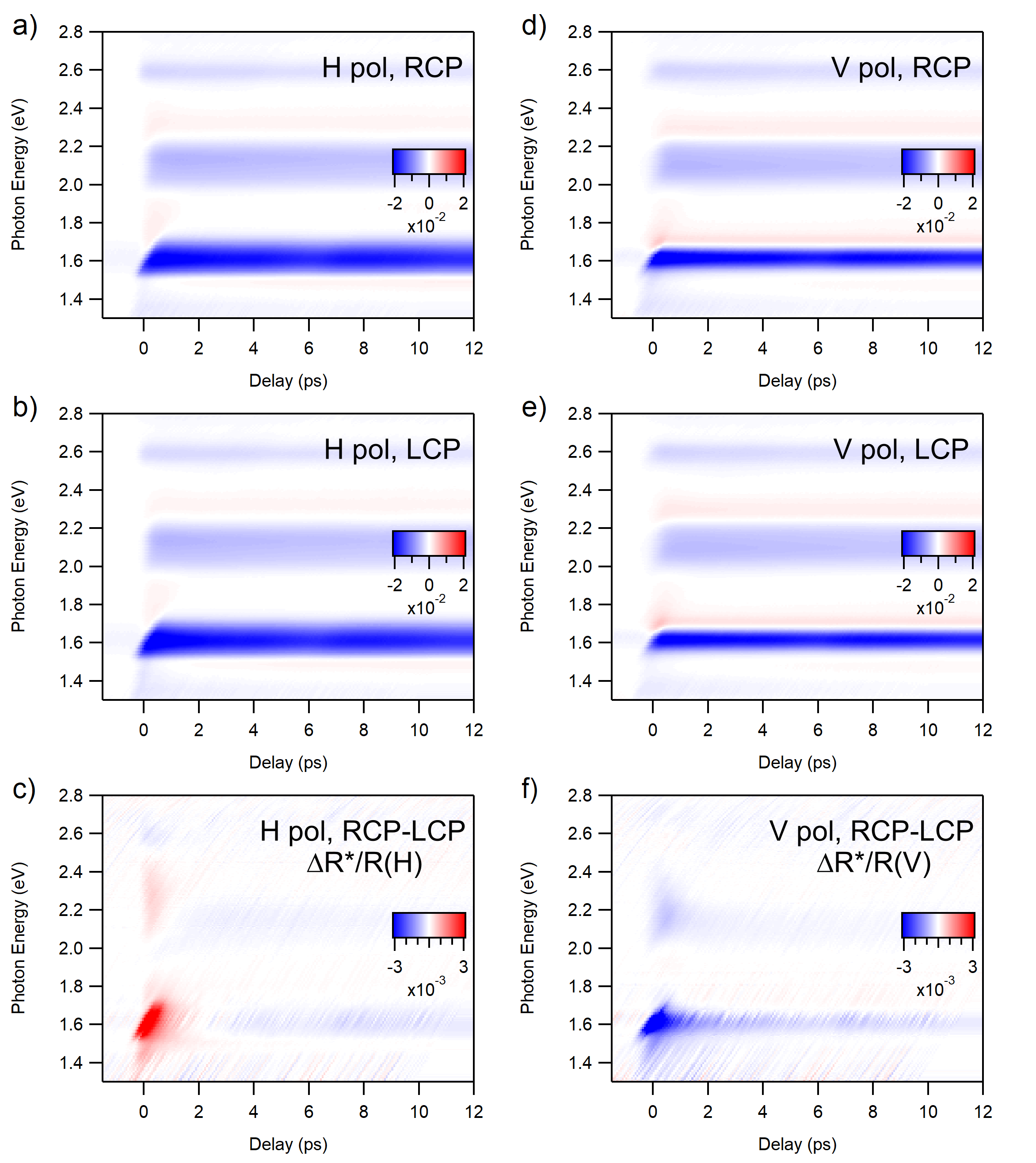}
\caption{
Differential reflectivity maps $\Delta R/R$ for bulk 2H-WSe$_2$
with pump photon energy at $\omega_{\rm pump}=1.55$ eV.
Panels (a)-(b) show $\Delta R/R$
along the H direction for RCP and LCP, respectively,
while panel (c) shows the dichroism $\Delta R^*/R(=\Delta R/R_{\rm RCP}-\Delta R/R_{\rm LCP}$.
Panels (d)-(f): as in panel (a)-(c) but along the V direction.
The remarkable dichroism in both the H and V directions points out the
evidence of a relevant Kerr rotation.
}
\end{figure}

\newpage

\begin{figure}
\includegraphics[width=13.5cm ]{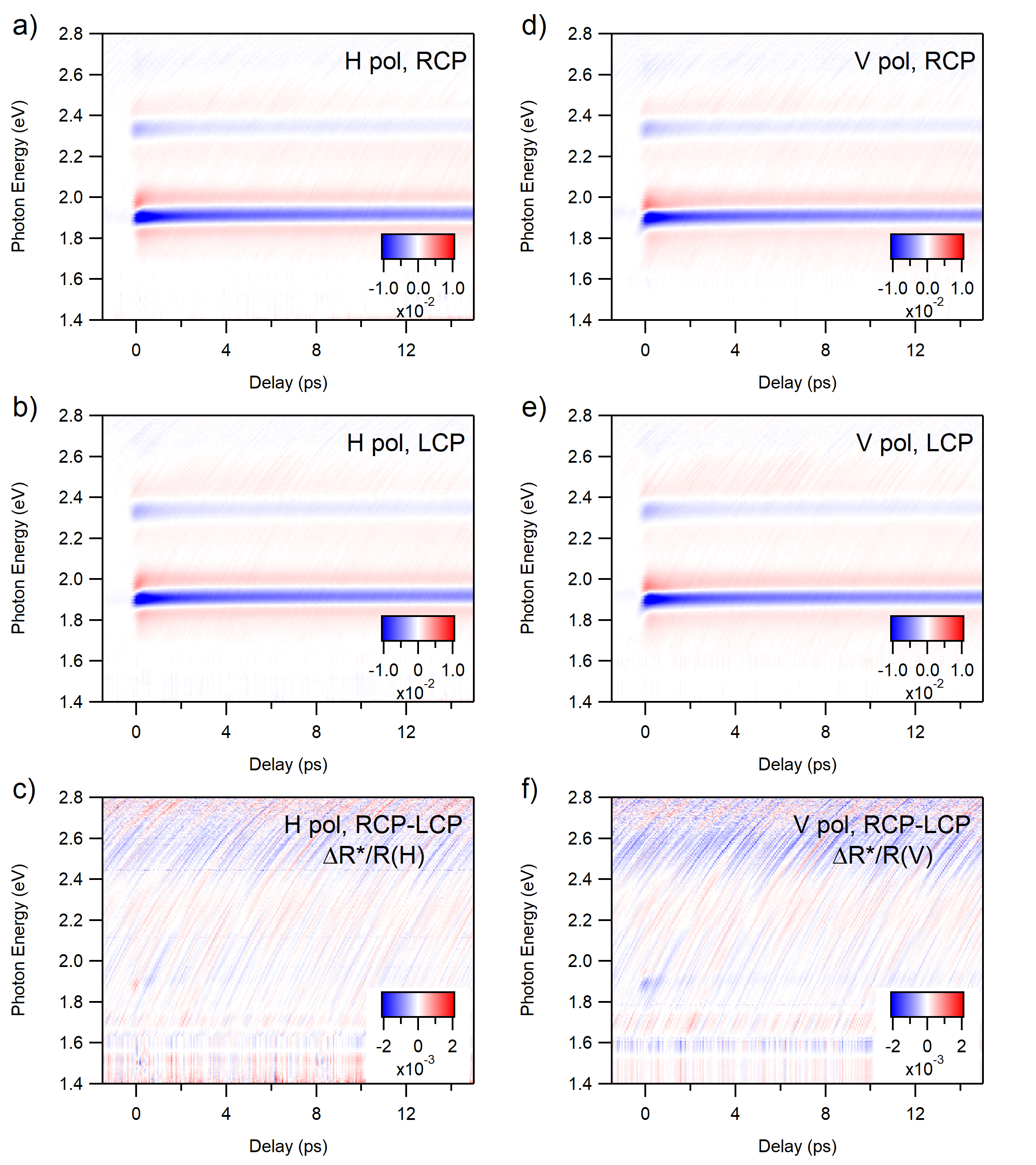}
\caption{
Differential reflectivity maps $\Delta R/R$ for bulk 2H-WS$_2$
with pump photon energy at $\omega_{\rm pump}=1.55$ eV.
Panels (a)-(b) show $\Delta R/R$
along the H direction for RCP and LCP, respectively,
while panel (c) shows the dichroism $\Delta R^*/R=\Delta R/R_{\rm RCP}-\Delta R/R_{\rm LCP}$.
Panels (d)-(f): as in panel (a)-(c) but along the V direction.
The lack of  dichroism in both the H and V directions point out the
absence of Kerr rotation under these circumstances.
}
\end{figure}

\newpage

\begin{figure}
\includegraphics[width=13.5cm ]{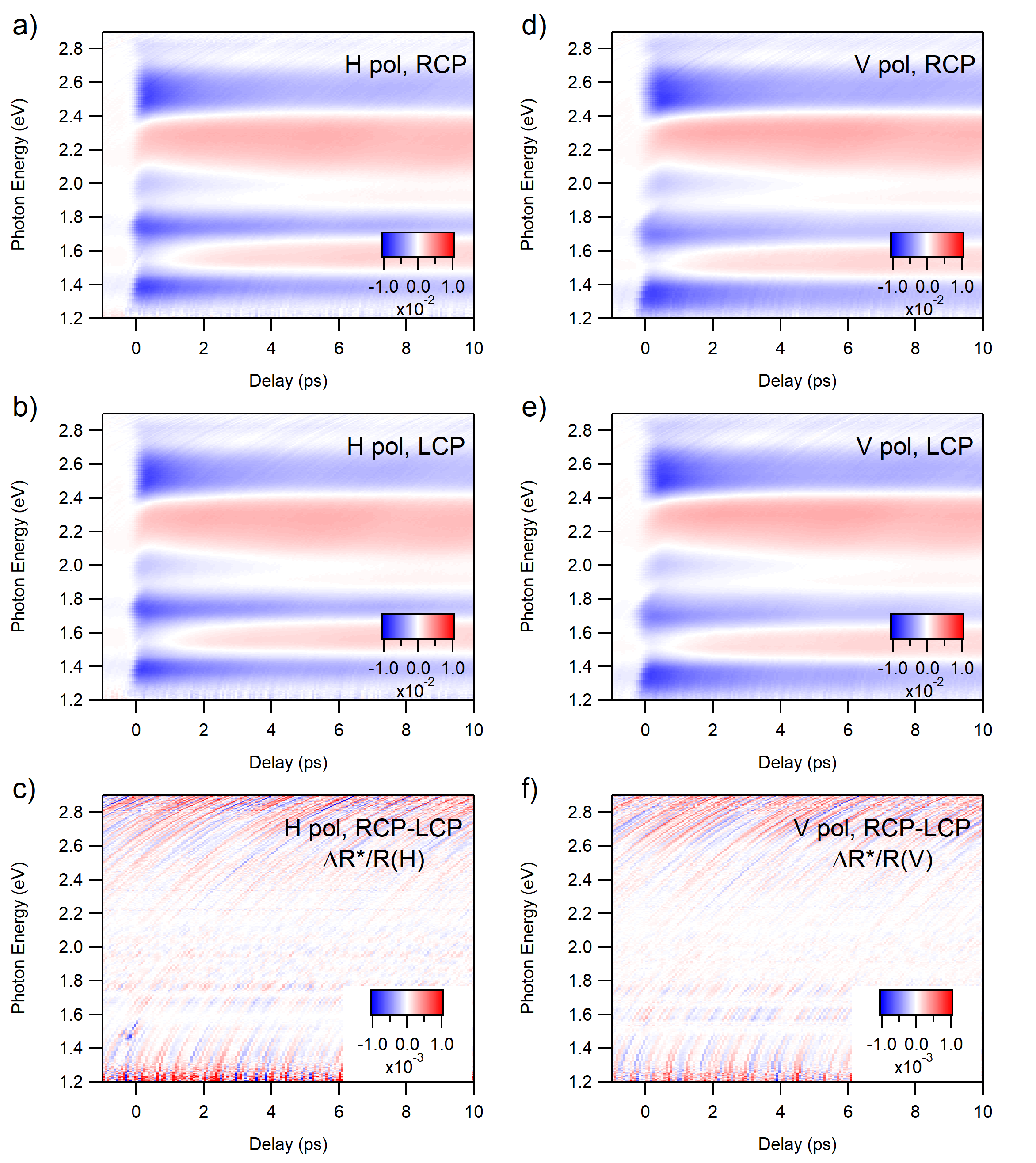}
\caption{
Differential reflectivity maps $\Delta R/R$ for bulk 2H-MoTe$_2$
with pump phonon energy at $\omega_{\rm pump}=1.55$ eV.
Panels (a)-(b) show $\Delta R/R$
along the H direction for RCP and LCP, respectively,
while panel (c) shows the dichroism $\Delta R^*/R=\Delta R/R_{\rm RCP}-\Delta R/R_{\rm LCP}$.
The lack of  dichroism in both the H and V directions point out the
absence of Kerr rotation under these circumstances.
}
\end{figure}